\documentclass[11pt,a4paper]{article}

\usepackage[utf8]{inputenc}
\usepackage{mathrsfs,amsmath} 
\usepackage{systeme}
\usepackage{mathtools}
\usepackage{authoraftertitle}
\usepackage{physics}
\usepackage{hyperref}
\usepackage{url}
\usepackage{subcaption} 
\usepackage{lipsum}
\usepackage[draft]{todonotes} 
\usepackage{chngcntr}
\usepackage{tikz-cd}
\usepackage{chemfig}
\usepackage{needspace}
\usepackage{cleveref}
\usepackage[absolute]{textpos}
\usepackage{dirtytalk}

\usepackage{mathrsfs,amsmath}

\usepackage[section]{placeins}
\usepackage{subfiles}

\usepackage[math]{cellspace}

\usepackage{graphicx,epsf}
\usepackage{array,multirow}
\usepackage{bm}
\usepackage{comment}

\renewcommand{\vec}[1]{\boldsymbol{#1}}

\newcommand{\ySubFig}[1]{
	\begin{subfigure}[t]
		{0.49\textwidth}
		\includegraphics[width=\textwidth]{./#1}
	\end{subfigure}
}

\newcommand{\yFigTwo}[4]{
	\begin{figure}[!ht]
		\ySubFig{#1}\ySubFig{#2}\caption{#3 \label{#4}}
	\end{figure}
}

\newcommand{\yFigFour}[6]{
	\begin{figure}[!ht]
		\ySubFig{#1}\ySubFig{#2}\ySubFig{#3}\ySubFig{#4}\caption{#5 \label{#6}}
	\end{figure}
}

\begin{document}

\begin{center}
	{\large{\textbf{Gyrokinetic modelling of the Alfv\'en mode activity in ASDEX Upgrade with an isotropic slowing-down fast-particle distribution}}}\\
	\vspace{0.2 cm}
	\author{F. Vannini}
	{\underline{F. Vannini}$^{1}$, A. Biancalani$^{4,1}$, A. Bottino$^1$, T. Hayward-Schneider$^{1}$, P. Lauber$^1$, A. Mishchenko$^2$, E. Poli$^1$, B. Rettino$^1$, G. Vlad$^3$, X. Wang$^1$ and the ASDEX Upgrade team$^5$.}
	
	\vspace{0.2 cm}

{
$^1$Max-Planck-Institut f\"ur Plasmaphysik, 85748 Garching, Germany \\
$^2$Max-Planck-Institut f\"ur Plasmaphysik, 17491 Greifswald, Germany\\
$^3$ENEA, Fusion and Nuclear Safety Department, 00044 Frascati (Roma), Italy\\
$^4$ L\'eonard de Vinci P\^{o}le Universitaire, Research Center, Paris la D\'efense, France\\
$^5$ See author list of \textit{H. Meyer et al. 2019 Nucl. Fusion \textbf{59} 112014}\\
}
	\vspace{0.2 cm}
	\small{francesco.vannini@ ipp.mpg.de\\
		}
\end{center}

\begin{abstract}

In the present paper, the evolution of the Alfv\'en modes is studied in a realistic ASDEX Upgrade equilibrium by analyzing the results of simulations with the global, electromagnetic, gyrokinetic particle-in-cell code ORB5. The energetic particles are modelled both via the newly implemented isotropic slowing-down and with Maxwellian distribution functions. The comparison of the numerical results shows that modelling the energetic particles with the equivalent Maxwellian rather than with the slowing-down, does not affect the frequency of the driven Alfv\'en mode, while its growth rate appears to be underestimated with a quantitative difference as large as almost $30\,\%$. Additionally the choice of the isotropic slowing-down allows a better description of the nonlinear modification of the dominant Alfv\'en mode frequency, while an equivalent Maxwellian underestimates it. A good comparison with the experimental spectrogram is found.


\end{abstract}

\clearpage
\section{Introduction}\label{Sec:Introduction}
Next-generation fusion machines like ITER \cite{Tomabechi_1991} and DEMO \cite{Zohm_2013} will be characterized by a significant population of energetic particles (EPs). With this term we refer to fast ions in general that are going to be present in the confining machines as fusion products ($\alpha$-particles, $^{4}_{2}He$) or as products of auxiliary heating sources like neutral beam injection (NBI) or ion cyclotron resonance heating (ICRH). The typical EP velocity $v_{EP}$ is intermediate between the thermal bulk ions velocity $v_{th,i}$ and the thermal electron velocity $v_{th,e}$. EPs' gyroradius $\rho_{EP}$ is instead much larger than the bulk plasma species gyroradii ($\rho_{i}$ and $\rho_{e}$) \cite{Heidbrink1994}:
\begin{equation}
    v_{th,i}\ll v_{EP}\ll v_{th,e}\quad ,\quad \rho_{EP}\gg \rho_{i} \gg \rho_{e}\quad .
\end{equation}
Typically in Tokamak machines the EPs' characteristic dynamical frequencies $\omega$ associated with their guiding-center motion (transit, bounce and precessional) fall inside the magnetohydrodynamic (MHD) regime $\omega \sim 10^{-2}\,\omega_{ci}$, with $\omega_{ci}$ the ion cyclotron frequency. Because of this EPs can resonate with MHD instabilities driving them unstable. Among the MHD instabilities, the most detrimental and easily excited are the nearly incompressible shear Alfv\'en wave (SAWs). These are transverse electromagnetic waves that propagate along the magnetic field lines with group velocity equal to the Alfv\'en speed $v_{A}$. Their dispersion relation is \cite{Mahajan_1995,Vlad1999DynamicsOA}:
\begin{equation}
    \omega_{SAW} = k_{\parallel}v_{A}\quad ,\quad k_{\parallel}=\Vec{k}\cdot\hat{\vec{b}}\quad,\quad v_{A}=\frac{B_{0}}{\sqrt{4\pi\rho_{m,0}}}\quad ,
    \label{Eq:dispersion}
\end{equation}
with $\hat{\vec{b}}=\vec{B}_{0}/B_{0}$ a unit vector pointing in the direction of the background magnetic field $\vec{B}_{0}$ with modulus $B_{0}$, $\rho_{m,0}$  the plasma mass density and $\vec{k}$ the wave-number of the perturbation. In a straight cylinder \cite{BARSTON1964282,sedlacek_1971,grad1969plasmas}:
\begin{equation}
    k_{\parallel}=\frac{1}{R_{0}}\left(n-\frac{m}{q(r)}\right)\quad ,
    \label{Eq:SAW_cylindrical}
\end{equation}
with $q(r)$ the safety factor profile, $R_{0}$ the major radius of the Tokamak and $m$ and $n$ the poloidal and toroidal mode numbers respectively.
SAWs are characterized by a frequency spectrum that varies continuously across the radial domain because of the radial dependence contained in both the Alfv\'en speed and the safety factor profile.\\ 
Different kinds of Alfv\'en instabilities, that we call for simplicity Alfv\'en modes (AMs) \cite{pwl2013,Heidbrink2008,ChenZonca2016}, can be present in Tokamak machines. These include both energetic-particle continuum modes (EPMs) \cite{Chen_1994} and Alfv\'en eigenmodes (AEs) \cite{Heidbrink2008}.\\
EPMs represent forced  oscillations of the SAW continuum that arise when the EP pressure is comparable to that of the bulk plasma. They emerge as discrete oscillations at the frequency of the continuum where the wave-EP power exchange is maximized, above the threshold of the continuum damping.\\ 
The AEs, on the other hand, are normal modes of the bulk plasma that can be classified into two types. The first type includes modes that arise in correspondence of the radial position $r_{0}$ where different branches of the continuum, see eq.~(\ref{Eq:SAW_cylindrical}), cross. There two counterpropagating waves interfere destructively, opening a gap in the continuum where modes reside. An important example of gap modes is represented by the toroidal Alfv\'en eigenmodes (TAE \cite{CHENG198521,Cheng1986,Heidbrink1991,Vlad1999DynamicsOA}) that arise because of the coupling between modes with close poloidal harmonics: $(m_{0},n)$ and $(m_{0}+1,n)$. The second type of AEs arise in correspondence of an extremum of the continuum spectrum. There $\partial\omega/\partial r$ vanishes and an effective potential well is established that traps the wave. An example of these kind of modes is represented by the reversed shear Alfv\'en eigenmodes (RSAE) \cite{Sharapov20022027,Berk20011850021} that are present in correspondence of a minimum in the safety factor profile.\\
The interest connected with the study of the driven AMs relies on their potentially detrimental role. In fact the driven AMs can interact with the EPs redistributing them in phase-space,  enhancing the EP transport. In this way EPs can be expelled before they can thermalize with the bulk plasma leading to a less effective heating \cite{McGuire1983,White1983,Sigmar1992}. Additionally AMs are held responsible for the presence of the so-called abrupt-large-events (ALE \cite{Shinohara_2001,Shinohara_2004,Ishikawa_2005,Bierwage-2018}). These are violent and rapid variations of the fluctuating magnetic field in correspondence of which an intense migration of EPs from the core to the periphery is observed, representing a possible threat for the safety of the machine. Therefore, the study of the AM dynamics is of primary importance both for the safety of the machine and because only a small fraction of EP losses can be tolerated to achieve ignition.  That is why is mandatory to gain the necessary insight in the AM dynamics to become predictive about the scenarios that will be met in ITER and DEMO.\\ 
In this paper we want to contribute to this task studying the AM dynamics with the code ORB5 \cite{LANTI2020107072} in numerical simulations where the so-called \say{NLED-AUG case} \cite{Lauber} is considered. With this term we refer to the plasma conditions of the ASDEX Upgrade discharge $\#31213$ at time $t=0.84\,$s. The uniqueness of this scenario relies on the fact that it has been obtained tuning the plasma parameters so that the EPs injected through a neutral beam (NB) have an energy with respect to the bulk plasma temperature $\mathcal{E}_{EP}/T_{Bulk}\sim 10^{2}$ and an induced fast-ion $\beta$ comparable to that of the bulk plasma.
In such a way there is a strong Alfv\'en activity, the stabilizing effects of the bulk plasma being minimized.
In particular, TAE-EPM bursts are observed in correspondence of which energetic particle driven geodesic acoustic modes (EGAMs \cite{Qiu_2011}) appear exhibiting the typical chirping.\\
Studies of the AM activity with the numerical tool ORB5 \cite{LANTI2020107072} using the NLED-AUG scenario have already been presented in Ref.~\cite{VANNINI2020,VANNINI2021}. There the EPs have been modelled, respectively, with Maxwellian distribution functions and with bi-shifted Maxwellians (also known as \say{double-bump-on-tail}). These were the two kinds of equilibrium distribution functions available at that time in ORB5 and the choice of one over the other was motivated by the kind of physics we were interested to reproduce. In particular the double-bump-on-tail was chosen because of the need to have an anisotropy in velocity space to drive unstable an EGAM and study its interaction with the AMs. In the works already published a good qualitative comparison with the experiments has been obtained. In this work we want to make a further step showing that, through the slowing-down distribution function \cite{gaffey} newly implemented in ORB5 we are able to go closer to the experimental conditions obtaining an even better quantitative comparison with the experiment.\\
The present paper is structured as follows. In Sec.~\ref{Sec:Num_model} the main features and the model of ORB5 are described. In Sec.~\ref{Sec:NLED_scenario} the NLED-AUG case is briefly described. In Sec.~\ref{Sec:slowing_down}  an analytical derivation of the slowing-down distribution function together with details about its implementation in ORB5 are provided. In Sec.~\ref{Sec:NLED} the results of numerical simulations obtained taking into account the NLED-AUG case are described. Finally in Sec.~\ref{Sec:conclusion} the conclusions of this paper are presented.

\section{The numerical model}\label{Sec:Num_model}
ORB5 \cite{LANTI2020107072} is a nonlinear, global, electromagnetic, gyrokinetic, particle-in-cell (PIC) code that can take into account collisions and sources (neglected in this work).\\ 
The code uses a system of straight-field line coordinates: $(r,\theta^{*},\varphi)$. As radial coordinate the code takes into account  the square root of the poloidal flux $\psi$ normalized at its value at the edge $\psi_{0}$: $r=\sqrt{\psi/\psi_{0}}$ with $0\leq r\leq  1$ that labels the magnetic surfaces. $\varphi$ is the toroidal angular coordinate and $\theta^{*}$ is the poloidal magnetic angle: 
\begin{equation}
    \theta^{*}=\frac{1}{2\,q(r)}\int_{0}^{2\pi}d\theta^{'}\frac{\vec{B}_{0}\cdot\nabla\varphi}{\vec{B}_{0}\cdot\nabla\theta^{'}}\quad ,\quad 0\leq\theta^{*}\leq2\pi
\end{equation}
with $\theta^{'}$ the geometric poloidal angle and $q(r)$ the safety factor profile.\\ 
In ORB5 all the physical quantities are normalized with respect to reference parameters. The masses of the considered particle species $m_{sp}$ are given in units of the mass of the main bulk ion species $m_{i}$, the velocities are given in units of the sound velocity $c_{s}$, the lengths are given in units of the sound Larmor radius $\rho_{s}$ and the time is given in units of the inverse bulk ion cyclotron frequency: 
\begin{equation}
    \omega_{ci}=\frac{q_{i}B_{0}}{m_{i}c}\quad ,\quad q_{i}=e\,Z_{i}\quad 
\end{equation}
with $e$ the electron charge in absolute value and $Z_{i}$ the atomic number of the main ion species.\\ 
The total distribution function of the $sp$-particle species $f_{sp}$ is divided into a time-independent part (equilibrium or background distribution function, $F_{0,sp}$) and a time dependent component $\delta f_{sp}$, so that $f_{sp}=F_{0,sp}+\epsilon_{\delta}\delta f_{sp}$ with $\epsilon_{\delta}$ a small parameter. Only the time-dependent component is discretized as will be later discussed.  The \textit{gyrokinetic Vlasov equation} for the perturbed (time-dependent) distribution function is:
\begin{equation}
 \frac{d }{dt}\delta f_{sp}=-\dot{\Vec{X}}\cdot\frac{\partial F_{0,sp}}{\partial \Vec{X}}\bigg\rvert_{\mathcal{E},\,v_{\parallel}} - \dot{\mathcal{E}}\frac{\partial F_{0,sp}}{\partial \mathcal{E}}\bigg\rvert_{\vec{X},\,v_{\parallel}}
    - \dot{v}_{\parallel}\frac{\partial F_{0,sp}}{\partial v_{\parallel}}\bigg\rvert_{\vec{X},\,\mathcal{E}}
    \quad ,\quad  \mathcal{E}=\frac{v_{\parallel}^{2}}{2}+\mu B \quad ,\quad  \mu=\frac{v_{\perp}^{2}}{2 B}
    \label{EqC2:Vlasov}
\end{equation}
where $\vec{X}$ is the gyrocenter position, while $v_{\parallel}$ and $v_{\perp}$ are, respectively, the parallel and perpendicular components of the particle velocity.
The equation of motions in mixed-variable formulation \cite{ami_2014} of the \textit{gyrocenter characteristics} of the particle species are \cite{ami_2019}:
\begin{multline}
    \Vec{\dot{X}} =v_{\parallel}\hat{\vec{b}}^{*}+\frac{1}{q_{sp}B^{*}_{\parallel}}\hat{\vec{b}}\times \mu\nabla B+\epsilon_{\delta}\left[
    \frac{\hat{\Vec{b}}}{B_{\parallel}^{*}}\cross\nabla \langle \delta\phi-v_{\parallel}\delta A_{\parallel}^{h}-v_{\parallel}\delta A_{\parallel}^{s}\rangle_{\alpha}-\frac{q_{sp}}{m_{sp}}\langle \delta A_{\parallel}^{h}\rangle_{\alpha}\hat{\Vec{b}}^{*}\right]
    \label{Eq:ORB1}
\end{multline}
\begin{multline}
    \dot{v}_{\parallel}=-\frac{\mu}{m_{sp}}\hat{\vec{b}}^{*}\cdot B
    -\epsilon_{\delta}\bigg\{\mu\frac{\Vec{\hat{b}}\cross \nabla B}{B_{\parallel}^{*}}\cdot \nabla\langle \delta A_{\parallel}^{s} \rangle_{\alpha}
    +\frac{q_{sp}}{m_{sp}}\left[\hat{\Vec{b}}^{*}\cdot\nabla\langle \delta\phi -v_{\parallel}\delta A_{\parallel}^{h}\rangle_{\alpha}+\frac{\partial }{\partial t}\langle \delta A_{\parallel}^{s}\rangle_{\alpha} \right]\bigg\}
    \label{Eq:ORB2}
\end{multline}
\begin{equation}
    \dot{\mathcal{E}}=v_{\parallel}\dot{v}_{\parallel}+\mu\nabla B\cdot \dot{\Vec{X}} \quad ,\quad \dot{\mu}=0\quad .
     \label{Eq:ORB3}
\end{equation}
In \cref{Eq:ORB1,Eq:ORB2,Eq:ORB3}, $m_{sp}$ and $q_{sp}=e\,Z_{sp}$ are respectively the mass and charge of the $sp$-particle species with atomic number $Z_{sp}$. Note that for the electron species $Z_{e}=-1$. 
Still in  \cref{Eq:ORB1,Eq:ORB2,Eq:ORB3} the gyroaveraging operator appears:
\begin{equation}
    \langle F\rangle_{\alpha}=\frac{1}{2\pi}\int_{0}^{2\pi} d\alpha\,F(\vec{X}+\vec{\rho}_{0})=\frac{1}{2\pi}\int_{0}^{2\pi} d\alpha\int d^{3}\vec{r}\,F(\vec{r})\,\delta^{3}(\vec{X}+\vec{\rho}_{0}-\vec{r})=\mathcal{J}^{gc}_{0}(F)
\end{equation}
that removes the fast gyroangle $\alpha$ dependence into a general quantity $F$, with $\vec{\rho}_{0}(\vec{X},\alpha)$ the gyroradius. $\delta\phi$ is the perturbed scalar potential and $\delta A_{\parallel}$ is the perturbed magnetic parallel potential decomposed into symplectic and Hamiltonian parts: $\delta A_{\parallel}=\delta A_{\parallel}^{h}+\delta A_{\parallel}^{s}$.
$B^{*}_{\parallel}$ is the parallel component of the symplectic magnetic field $\vec{B}^{*}$, linked to the symplectic magnetic potential $\vec{A}^{*}$ by the following relation:
\begin{equation}
    \vec{A}^{*}=\vec{A}+\frac{m_{sp}}{q_{sp}}v_{\parallel}\hat{\vec{b}}\quad ,\quad \vec{B}^{*}=\nabla\times \vec{A}^{*}\quad ,\quad \hat{\vec{b}}^{*}=\frac{\vec{B}^{*}}{B^{*}_{\parallel}}= \frac{\vec{B}^{*}}{\hat{\vec{b}}\cdot\vec{B}^{*}}\quad .
    \label{eq:A*}
\end{equation}
In ORB5 only the perpendicular component of the perturbed magnetic field $\delta\vec{B}_{\perp}$ is implemented: $\delta\vec{B}_{\perp}=\hat{\vec{b}}\times\nabla\delta A_{\parallel}$. In eq.~(\ref{eq:A*}) $\vec{A}$ is the background magnetic potential: $\vec{B}_{0}=\nabla\times\vec{A}$ and $\hat{\vec{b}}$ a unit vector pointing in the direction of the background magnetic field. The characteristic equations \cref{Eq:ORB1,Eq:ORB2,Eq:ORB3} are coupled to the field equations. These are the \textit{gyrokinetic quasi-neutrality equation}:
\begin{equation}
-\nabla\cdot \left[\left(\sum_{sp=i,f}\frac{q_{sp}^{2}n_{sp}}{T_{sp}}\rho_{sp}^{2}\right)\nabla_{\perp}\delta\phi\right]=\sum_{sp=i,e,f}q_{sp}\,\delta n_{sp}\quad ,\quad \delta n_{sp}= \int dW \langle \delta f_{sp}\rangle   
\label{Eq:quasineutrality}
\end{equation}
the \textit{parallel Ampère’s law}
\begin{equation}
    \left(\sum_{sp=i,e,f}\frac{\beta_{sp}}{\rho_{sp}^{2}}-\nabla_{\perp}^{2}\right)\delta A_{\parallel}^{h}=\mu_{0}\sum_{sp=i,e,f}\delta j_{\parallel,sp}+\nabla^{2}_{\perp}\delta A_{\parallel}^{s} \quad ,\quad  \delta j_{\parallel,sp}=q_{sp}\int dW \,v_{\parallel}\,\langle \delta f_{s}\rangle
    \label{Eq:Ampere}
\end{equation}
and the \textit{ideal Ohm’s law}:
\begin{equation}
    \frac{\partial }{\partial t}\delta A_{\parallel}^{s}+\hat{\Vec{b}}\cdot \nabla\delta \phi = 0
    \label{Eq:IdealOhm}
\end{equation}
where:
\begin{equation}
    n_{sp}=\int dW F_{0,sp}\quad ,\quad  \beta_{sp}=\mu_{0}\frac{n_{sp}T_{sp}}{B_{0}^{2}}
\end{equation}
where $dW=B^{*}_{\parallel}\,d v_{\parallel}\,d\mu\, d\alpha$.
\Cref{EqC2:Vlasov,Eq:ORB1,Eq:ORB2,Eq:ORB3,Eq:quasineutrality,Eq:Ampere,Eq:IdealOhm}  constitute  the gyrokinetic Vlasov-Maxwell system of equations solved by ORB5. \\
The discretization is achieved sampling the phase-space with a set of super-particles called \textit{markers}. The $k$-th
 marker of the $sp$-particle species is associated to a weight $\omega_{sp,k}(t)$. 
 This is a time-dependent quantity that represents at the time $t$ the variation of the number of physical particles contained
in a small volume in phase-space $\Omega_{k}$, centered around the $k$-th marker:
 \begin{equation}
    \delta N_{sp}^{physical\,particles}=\int_{\Omega_{k}}\,d^{6}\vec{Z}\,\delta f_{sp}(\vec{Z}(t),t)=\omega_{sp,k}(t)\quad .
\end{equation}
Each marker is pushed along its orbit slowing the characteristics equations through a Runge-Kutta method at $4$th order.\\ 
The perturbed fields $\Psi=\{\delta\phi,\delta A_{\parallel}\}$ are discretized using the finite-elements Galerkin approximation. The fields are represented as linear combinations of finite dimensional function space $\Lambda_{\mu}(\vec{x})$:
\begin{equation}
    \Psi(\vec{x},t)=\sum_{\mu}\Psi_{\mu}(t)\Lambda_{\mu}(\vec{x})\quad 
    \label{Eq:discr_fields}
\end{equation}
where the basis $\Lambda_{\mu}(\vec{x})$ are tensor product of $1$D polynomials (B-splines) of degree $p=1,2,3$:
\begin{equation}
    \Lambda_{\mu}(\vec{x})=\Lambda_{j}^{p}(r)\Lambda_{k}^{p}(\theta^{*})\Lambda_{l}^{p}(\varphi)\quad \text{with}\quad \mu=(j,k,l)\quad .
\end{equation}
In the present work $p=3$. 
The number of B-splines in each direction is connected to the number of grid points (knots) $(N_{r},N_{\theta^{*}},N_{\varphi})$ associated to the B-splines and to the chosen degree $p$ of the polynomials. Through the decomposition expressed in eq.~(\ref{Eq:discr_fields}) the fields equations become a set of linear equations to which is applied the double discrete Fourier transformation $\mathcal{F}$ in both the poloidal and toroidal directions. Inverting the obtained system of equations, it is then possible to calculate the Fourier coefficients of the perturbations, to which a Fourier filter is applied. This is given by a field aligned filter:
\begin{equation}
    m\in \underbrace{[n\,q(r)-\Delta m,n\,q(r)+\Delta m]}_{\text{field aligned filter}}\quad ,\quad n\in[nfilt1,nfilt2]\quad .
    \label{EqC2:filter}
\end{equation}
In eq.~(\ref{EqC2:filter}) $nfilt1$, $nfilt2$ are the range boundaries of the rectangular filter for the toroidal coefficients. $\Delta m$, instead, is the width of the field aligned filter.\\
In Sec.~\ref{Sec:NLED} results of ORB5 simulations will be discussed. There we will indicate some important parameters used in the simulations. We will specify: the number of grid points associated to the B-splines $(N_{r},N_{\theta^{*}},N_{\varphi})$, the number of markers used for every particle species $nptot_{sp}$, the time step in use $\Delta t$ and the width of the field aligned filter $\Delta\,m$. In this work only the AMs in the NLED-AUG scenario are investigated by retaining, as in previous works \cite{VANNINI2020,Vlad2021}, only the modes with toroidal mode number equal to one: $nfilt1=nfilt2=1$. Additionally only the EPs will be allowed to redistribute in phase-space, following their full trajectories. The bulk plasma species instead will follow their unperturbed trajectories. This means that in \cref{Eq:ORB1,Eq:ORB2,Eq:ORB3} the small parameter is taken, with abuse of notation, equal to one only for the EPs while for the electrons and the bulk ions species it is identically zero.\\
In the following sections we will also specify if a drift-kinetic model is considered and finite-Larmor-radius (FLR) are neglected, or if a gyrokinetic model is considered and FLR effects are taken into account.\\
\section{NLED-AUG scenario}\label{Sec:NLED_scenario}
\begin{figure}[!h]
    \centering
    \includegraphics[width=0.9\textwidth]{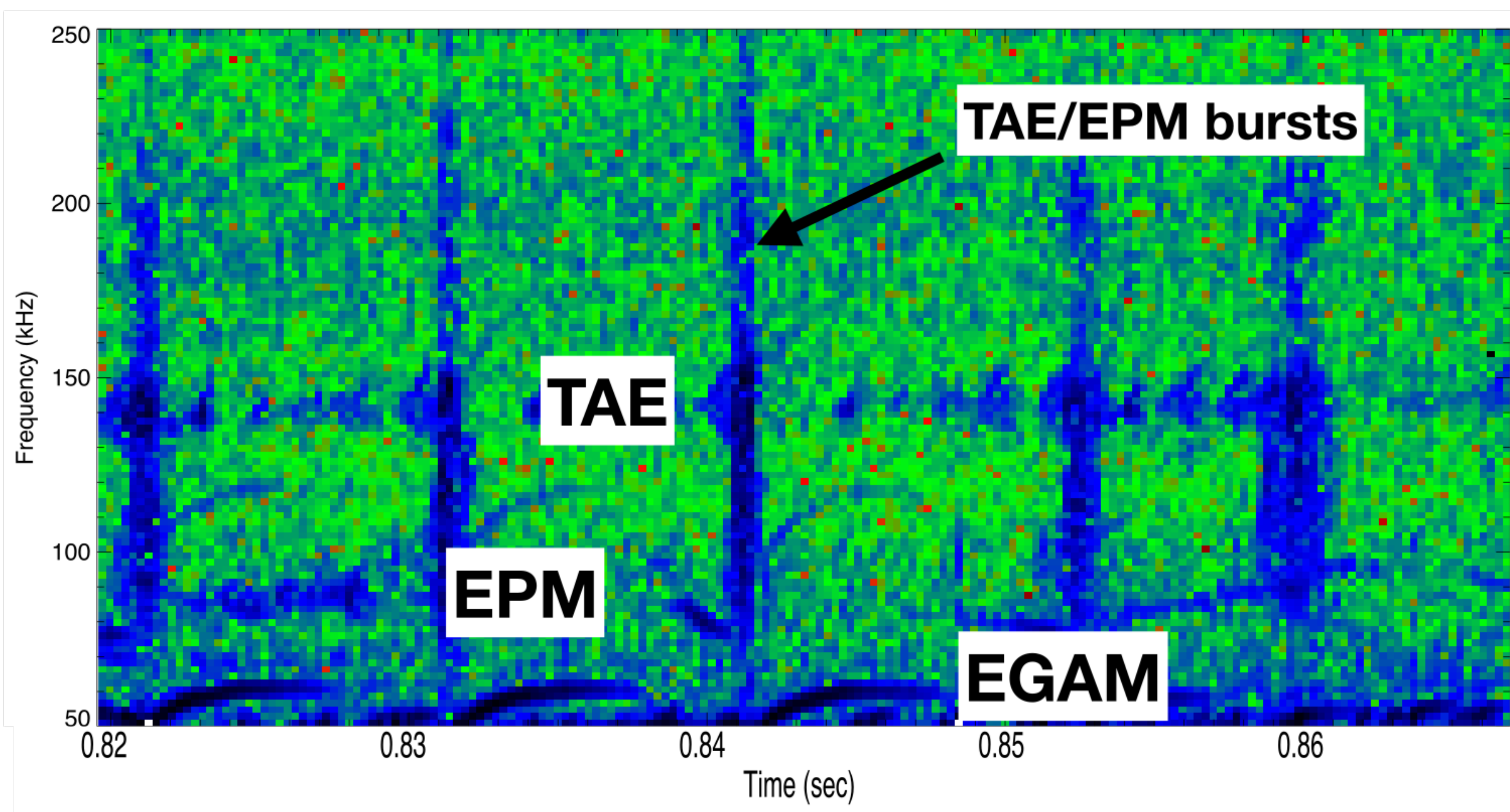}
    \caption{Experimental spectrogram. At $t=0.84\,$s the plasma parameters, profiles and the reconstructed magnetic equilibrium have been selected to perform numerical simulations. At this time a TAE-EPM burst, covering the frequency range $80\,$kHz$\,\leq\nu\leq250\,$kHz, is observed. At lower frequencies $\nu\approx 50\,$kHz, an EGAM exhibits the typical chirping behaviour.}
    \label{Fig:experimental_spectrogram}
\end{figure}

As stated in the introduction, with the term NLED-AUG case \cite{Lauber2} we refer to the plasma conditions present at $t=0.84\,$s in the discharge number $31213$  ($\#31213@0.84$s) performed in the Tokamak ASDEX Upgrade (AUG). EPs, in this experimental case, are due to a NB launched with an injection angle of $7.13^{\circ}$ with respect to the horizontal plane. This experimental case exhibits, as anticipated in Sec.~\ref{Sec:Introduction}, plasma parameters previously unexplored in ASDEX Upgrade. In fact, the NB-induced fast-ion $\beta$ is comparable to that of the bulk plasma and EPs' injection energy is approximately $100$ times higher than the bulk species temperatures:
\begin{equation}
 \beta_{EP}/\beta_{Bulk}\sim 1\quad ,\quad \mathcal{E}_{EP}/T_{Bulk}\approx 93\,keV/1\,keV\sim 10^{2}\quad .
    \label{EqC4:NLED_parameters}
\end{equation}
By doing so, the realistic ratios of plasma parameters that are going to be met in future fusion machines have been achieved in this scenario (in ITER/DEMO:  $\mathcal{E}_{EP}/T_{Bulk}\approx 3.5 MeV/30keV$). In such a way an intense EP-driven activity is observed, with the stabilizing effects of the bulk plasma minimized. That is why this is a very interesting scenario to understand the EP-driven dynamics and to validate the codes against the experiments. In particular  this case presents a rich nonlinear physics of interest. Around $t\approx 0.84\,$s a TAE-EPM burst is present after which, at lower frequencies, an EGAM exhibit the typical chirping behaviour (cf. Fig.~\ref{Fig:experimental_spectrogram}). According to the pick-up-coils measurements, in this scenario the most unstable AM has toroidal mode number $n=1$.
\\
In Fig.~\ref{FigC4:NLEDplasmaprofiles} the temperature profiles of the bulk plasma species (left) and the radial dependence of the electron density profile (right) are shown, while 
\yFigTwo{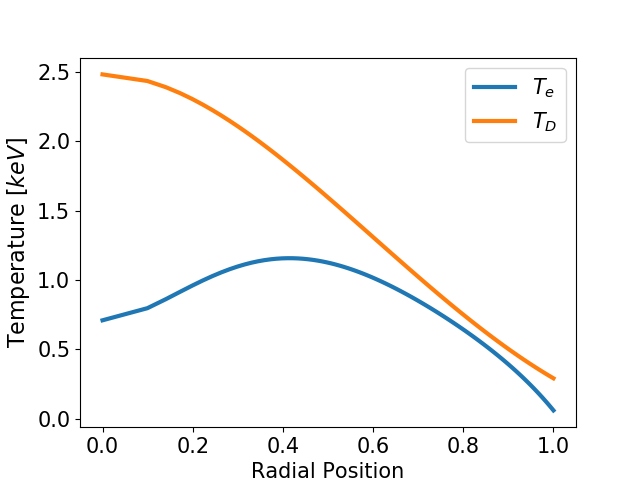}{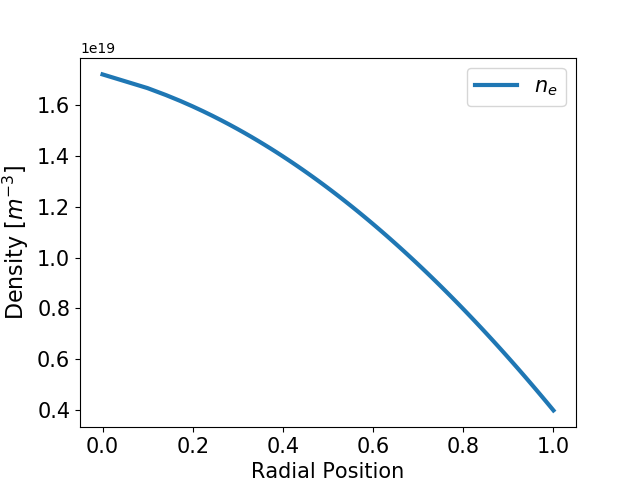}{Radial temperature profiles (\textbf{left}) of the bulk plasma species (electrons and deuterium) and radial electron density profile (\textbf{right}) of the NLED-AUG case.
}{FigC4:NLEDplasmaprofiles}
in Fig.~\ref{FigC4:NLEDqprofile} left the radial dependence of the EP density profile $n_{EP}$ modelled by TRANSP \cite{hawryluk1981empirical} is shown. The EPs have an off-axis radial density profile with a reference  concentration of $\langle n_{EP}\rangle/\langle n_{e}\rangle=0.0949$, where $\langle ...\rangle$ indicates volume average.
In Fig.~\ref{FigC4:NLEDqprofile} right the radial dependence of the safety factor profile $q$ is shown. It has a reversed shear, with a minimum around $r\approx 0.5$. Additionally the TAE position $r\approx 0.738$ is indicated (black point). The TAE is formed in the gap created by the interaction between the two most unstable scalar potential Fourier components: $(m,n)=(2,1)$ and $(m,n)=(3,1)$.
\yFigTwo{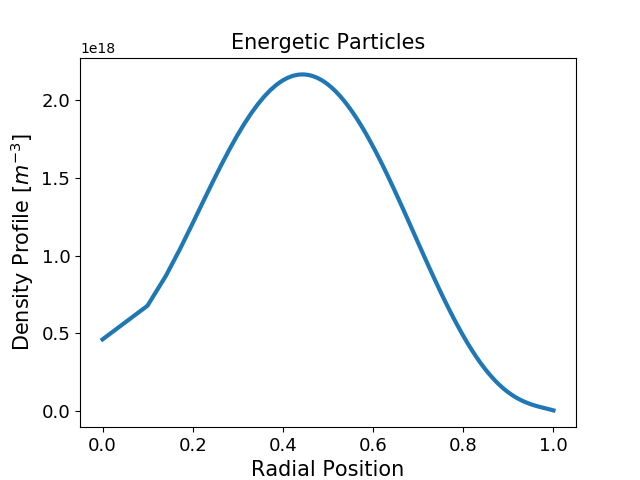}{figs/NLED_AUG/q_NLED.png}{\textbf{Left:} EPs off-axis radial density profile as modelled by TRANSP. It  has a reference concentration of $\langle n_{EP}\rangle/\langle n_{e}\rangle=0.0949$.  \textbf{Right:} Safety factor profile of the NLED-AUG case. It exhibits  a non-monotonic profile with a minimum around $r\approx 0.5$ where $q\approx 2.28$ and a maximum at $r=1$ in the amount of $q\approx 8.5$. The inset figure marks the TAE location at $r_{0}\approx 0.738$, corresponding to $q_{0}=q(r_{0})=(2\, m_{0}+1)/2=2.5$, with $n=1$ and $m_{0}=2$.}{FigC4:NLEDqprofile}
\\
In Tab.~\ref{TabC4:NLED0} the values of some important constants that will be used in the simulations are shown. In this work frequencies and growth rates will be provided in $\omega_{A0}$-units, that is the Alfv\'en frequency on-axis, at $r=0$: $\omega_{A0}=v_{A}(r=0)/R_{0}$, with $v_{A}$ the Alfv\'en speed (cf. \cref{Eq:dispersion}) and $R_{0}$ the major radius of the tokamak.
\begin{table}[h!]
    \centering
    \begin{tabular}{||c|c|c|c|c|c|c|c||}
    \hline
        $a_{0}\,[$m$]$ &  $R_{0}\,[$m$]$ & $B_{0}\,[$T$]$&$\beta_{e}$ &$L_{x}$ &$\omega_{ci}\,[$rad/s$]$ & $\omega_{A0}\,[$rad/s$]$&$\omega_{ci}/\omega_{A0}$\\
        \hline
         $0.482$ &  $1.666$ & $2.202$& $2.7\cdot 10^{-4}$ &$551.6$& $1.055\cdot 10^{8}$ & $4.98\cdot 10^{6}$ & $\approx 21$\\
         \hline
    \end{tabular}
    \caption{Constants in use: averaged minor radius $a_{0}$, major radius $R_{0}$, amplitude of the background magnetic field on axis $B_{0}$. Normalized electron pressure $\beta_{e}=4\pi\, n_{e}(r_{0})T_{e}(r_{0})/B_{0}^{2}$ with the $r_{0}$ a reference radial position. Normalized size of the plasma system that is $L_{x}=2\,a_{0}/\rho_{s}$, with $\rho_{s}=c_{s}/\omega_{ci}$ and $c_{s}$ the sound velocity and $\omega_{ci}$ the ion cyclotron frequency. $\omega_{A0}$ is the value of the Alfv\'en frequency on axis.}
    \label{TabC4:NLED0}
\end{table}
\\
Many works have already been dedicated to the investigation of the mode dynamics of this interesting experimental case. The linear AM dynamics has been investigated in Ref.~\cite{VANNINI2020,Vlad2021} in simulations where both on-axis and off-axis radial density profiles have been considered for the EPs. These have been modelled via Maxwellian distribution functions associated to temperature profiles constant against the radius. In Ref.~\cite{NOVIKAU2020} numerical studies of the nonlinear EGAM dynamics have been conducted. In Ref.~\cite{Poloskei} bicoherence studies have suggested that nonlinear coupling exists between the observed TAE-EPM burst triggering the EGAM. In Ref.~\cite{VANNINI2021} the interaction between AMs and EGAMs has been studied both with numerical simulations and through an analytical model where only the EPs have been allowed to redistribute in phase-space, following their full trajectories. Additionally, there the EPs have been modelled via a double-bump-on-tail distribution function, since an anisotropy in velocity space was needed to drive an EGAM unstable \cite{Fu2008,Zonca_2008,Zarzoso2014}.\\
In the present work we focus on the AM dynamics, modelling the bulk plasma species with Maxwellian distribution functions and retaining the nonlinearities only in the EP dynamics, as in \cite{VANNINI2021,Vlad2021}. That is, only the EPs gyrocenter characteristics \cref{Eq:ORB1,Eq:ORB2,Eq:ORB3} have $\epsilon_{\delta}=1$. The other particle species (electrons and bulk ions) follow their unperturbed trajectories and for them $\epsilon_{\delta}=0$. The novelty of this work is represented by the fact that we go closer to the experimental conditions modelling the EPs via an isotropic slowing-down distribution function. We emphasize here that EPs are present in the NLED-AUG case as injected through an NB. This implies that their distribution function is intrinsically anisotropic in phase-space. In the studies carried out in this paper we neglect the effects of the anisotropy in velocity space that are, on the other hand, important to detail the interaction between AMs and EGAMs (that here we do not study in contrast to what was done in Ref.~\cite{VANNINI2021}). Nevertheless the isotropic slowing-down represents an improvement with respect to the Maxwellian distribution function, as will be motivated in Sec.~\ref{Sec:slowing_down}. Additionally through its choice we will observe a good quantitative comparison with the experiment making a further step with respect to previously published works where already a good qualitative agreement has been obtained. In the NLED-AUG scenario the bulk ions and the EPs are constituted by deuterium ions. In all the simulations discussed in this work the realistic electron mass will be considered: $m_{e}\approx m_{D}/3676$.\\
In Sec.~\ref{Sec:slowing_down} we report  a simplified analytical derivation and discuss why the slowing-down distribution function represents a better description of the EPs displacement in phase-space and why this is an improvement with respect to the Maxwellian and the double-bump-on-tail distribution functions. Later we will describe its implementation in ORB5.\\
In all the simulations presented in Sec.~\ref{Sec:NLED} the quasi-neutrality condition  will always be satisfied $ n_{e}=Z_{D}\,n_{D}+Z_{EP}\,n_{EP}$ with $Z_{D}=Z_{EP}=1$,
by keeping fixed the electron radial density profile (cf. Fig.~\ref{FigC4:NLEDplasmaprofiles} right) and varying accordingly the EP and bulk deuterium profiles. 
\section{Slowing-down distribution function}\label{Sec:slowing_down}

We want to investigate the AM dynamics in the NLED-AUG case with ORB5. Approximations are present in the performed simulations. The main approximations here concern the choice of not retaining the bulk plasma nonlinearities and the choice for the equilibrium distribution function of the particle species. The importance of modelling the particle species with a proper equilibrium distribution function can be understood by looking at the gyrokinetic Vlasov equation \cref{EqC2:Vlasov}. In fact $\delta f_{sp}$ can only be evolved once the equilibrium distribution function and its derivatives are known. Obviously changing the equilibrium distribution function implies a modification of the time evolution of the perturbed distribution function, thus catching different aspects of physics.\\
While particles of the bulk plasma are usually well described by Maxwellian distribution functions (that from now on we denote with $F_{M}$), the question here is what is the shape of the distribution function that better represents the arrangements of the EPs in phase-space. For this, we choose a non-Maxwellian equilibrium distribution, namely the \say{slowing-down} \cite{Cordey1974,gaffey,Mila_Candy_2006,Angioni2008,adisi2018} that, from now on, we simply indicate with $F_{0}$.\\
To derive its form we summarize the main steps presented in Ref.~\cite{gaffey} where the analytical form of the slowing-down is derived. There the author studies how the fast-ions with mass $m_{EP}$ and charge $q_{EP}$ injected in plasma through a beam are displaced in phase-space because of the collisions with the particles of the bulk plasma. The starting point is represented by the Fokker-Planck equation \cite{PhysRev.107.1} for the EPs:
\begin{equation}
    \partial_{t}F_{0}+\Vec{v}\cdot\nabla_{\Vec{x}}F_{0}+\frac{q_{EP}}{m_{EP}}\left(\Vec{E}_{0}+\frac{\vec{v}}{c}\times\Vec{B}_{0}\right)\cdot\nabla_{\Vec{v}}F_{0}=C(F_{0})+S\quad .
    \label{EqC6:Fokker-Planck}
\end{equation}
In eq.~(\ref{EqC6:Fokker-Planck}) $S$ is the EPs source, while $C(F_{0})$  \cite{Thompson1964} is the collision operator, composed by the sum of two contributions
\begin{equation}
    C(F_{0})=C(F_{0},F_{M,e})+C(F_{0},F_{M,i})\quad .
    \label{Eq:collision_operator_total}
\end{equation}
$C(F_{0},F_{M,e})$ takes into account the effects of the collisions of the EPs with the electrons while $C(F_{0},F_{M,i})$ considers the collisions of the EPs with the bulk ions. 
Considering an EP beam axisymmetrically distributed around the background magnetic field and writing explicitly the analytical form of the collision operator, eq.~(\ref{EqC6:Fokker-Planck}) becomes \cite{gaffey}:
\begin{equation}
     \partial_{t}F_{0}=\frac{1}{\tau_{s}v^{3}}\Bigg\{v\frac{\partial}{\partial v}\left[\left(v^{3}+v_{c}^{3}\right)F_{0}\right]+Z_{2}\frac{v_{c}^{3}}{2}\frac{\partial}{\partial\xi}\left[\left(1-\xi^{2}\right)\frac{\partial F_{0}}{\partial\xi}\right]\Bigg\}+S(v,\xi)\quad .
     \label{EqC6:Fokker-Planck_0}
\end{equation}
In eq.~(\ref{EqC6:Fokker-Planck_0}) $v$ is the modulus of the beam particle velocity $v=|\vec{v}|$ and $\xi=v_{\parallel}/v=\cos\chi$, where $v_{\parallel}$ is the component of the particle velocity parallel to the background magnetic field and $\chi$ the pitch angle. $v_{c}$ is the \say{crossover velocity} or \say{slowing-down critical velocity}:
\begin{equation}
    v_{c}=\left(\frac{3\sqrt{\pi}}{4}\frac{m_{e}}{m_{EP}}Z_{1}\right)^{1/3}v_{th,e}\quad ,\quad  Z_{1}=\sum_{i={Bulk\,ions\,species}}\frac{n_{i}}{n_{e}}\frac{m_{EP}}{m_{i}}Z_{i}^{2}\quad 
    \label{EqC6:cross-over}
\end{equation}
$v_{th,e}$ is the electron thermal velocity and
\begin{equation}
    Z_{2}=\sum_{i={Bulk\,ions\,species}}\frac{n_{i}\,Z_{i}^{2}}{n_{e}Z_{1}}\quad .
    \label{EqC6:cross-over1}
\end{equation}
The crossover velocity is a property of the bulk plasma and represents a critical speed as for $v > v_{c}$ the electron drag dominates over the ion drag and the EPs are mainly decelerated by friction with the electrons. For $v < v_{c}$ the opposite trend is met and the EPs are mainly decelerated by friction with the bulk ions.
$\tau_{s}$ \cite{HelanderBook} is the slowing-down time that  describes the time after which a particle of the energetic ion beam is decelerated by collisions with the other particle species.\\
We take into account a monoenergetic EP beam. In this way the EP source in eq.~(\ref{EqC6:Fokker-Planck_0}) becomes:
\begin{equation}
    S(v,\xi)=\frac{S_{0}}{v^{2}}\delta(v-v_{EP})\delta(\xi-\xi_{EP})\quad .
    \label{EqC6:source}
\end{equation}
$S_{0}$ is the intensity of the EP source, $\xi_{EP}$ is the injection parallel velocity of the beam normalized to the modulus of the total velocity $v_{EP}=\sqrt{2\mathcal{E}_{EP}/m_{EP}}$
with $\mathcal{E}_{EP}$ the injection energy of the beam.\\ In Ref.~\cite{gaffey} the solution of eq.~(\ref{EqC6:Fokker-Planck_0}) is nicely derived. This is quite complex and contains the dependence in $\xi$ (or $\chi$, the pitch angle). Here, and in the results presented below, we disregard the $\xi$-dependence. It is then straightforward to obtain the steady state solution for an isotropic, monoenergetic, EP beam:
\begin{equation}
    F_{0}(v) = \frac{S_{0}\tau_{s}}{4\pi}\frac{\theta(v_{EP}-v)}{v_{c}^{3}+v^{3}}\quad .
    \label{EqC6:slowing1}
\end{equation}
This $\xi$-independent slowing-down distribution function is appropriate to describe fusion-born  products \cite{Angioni2008} since $\alpha$ particles are not born in any preferential direction (they are isotropic). On the other hand eq.~(\ref{EqC6:slowing1}) does not appropriately describe the arrangement in phase-space of the EPs belonging to an injected ion beam. In this case, in fact, the EPs are not isotropic. Eq.~(\ref{EqC6:slowing1}) represents nevertheless an improvement with respect to the Maxwellian distribution function and it is widely used in literature to model EPs \cite{Angioni2008}. In fact, through the Heaviside step function $\theta(v_{EP}-v)$,  it provides a constraint in the velocity distribution of the EPs present (born/injected) in plasma with $\mathcal{E}_{EP}$, since $0\leq v\leq v_{EP}$. Additionally its dependence in velocity-space is governed by the crossover velocity that, as previously stated, is a property of the bulk plasma.\\
By taking the zero and second order moments of eq.~(\ref{EqC6:slowing1}) we find \cite{Mila_Candy_2006}:
\begin{equation}
    \int d^{3}\Vec{v}\,F_{0}= S_{0}\tau_{s}I_{2}\left(\frac{v_{c}}{v_{EP}}\right)\quad ,\quad \int d^{3}\Vec{v}\,v^{2}F_{0}= S_{0}\tau_{s}I_{4}\left(\frac{v_{c}}{v_{EP}}\right)
    v_{EP}^{2}\quad ,
    \label{EqC6:moment}
\end{equation}
where:
\begin{equation}
    I_{n}(a)=\int_{0}^{1}dx\frac{x^{n}}{x^{3}+a^{3}}=
    \begin{cases}
    n=2\quad ,\quad  \frac{1}{3}\ln(1+a^{-3})
    \\
    \\
    n=4\quad ,\quad  \frac{1}{2}-a^{2}\Big\{\frac{1}{6}\ln\left(\frac{1-a+a^{2}}{(1+a)^{2}}\right)+\frac{1}{\sqrt{3}}\left[\tan^{-1}\left(\frac{2-a}{a\sqrt{3}}\right)+\frac{\pi}{6}\right]\Big\}\quad  .
    \end{cases}
\end{equation}
To ensure the zero order moment in \cref{EqC6:moment} to give the EPs' radial density profile $n_{EP}$, we impose:
\begin{equation}
    S_{0}=\frac{n_{EP}}{\tau_{s}\,I_{2}\left(\frac{v_{c}}{v_{EP}}\right)}\quad .
\end{equation}
With this choice it follows:
\begin{equation}
    F_{0} = \frac{n_{EP}}{\frac{4\pi}{3}\log\left[1+\left(\frac{v_{EP}}{v_{c}}\right)^{3}\right]}\frac{\theta(v_{EP}-v)}{v_{c}^{3}+v^{3}} \quad 
    \label{EqC6:slowing2}
\end{equation}
that represents the analytical form of the slowing-down distribution function properly normalized. \Cref{EqC6:slowing2} has been implemented in ORB5 together with its derivatives that are needed to evolve the perturbed EP distribution functions $\delta f_{EP}$ \cref{EqC2:Vlasov}:
\begin{equation}
    \begin{cases}
    \frac{d\,F_{0}}{d\,\psi}=\Bigg\{\frac{d}{d\,\psi}\log n_{EP}+\left(\frac{(v_{EP}/v_{c})^{3}}{\left[1+\left(\frac{v_{EP}}{v_{c}}\right)^{3}\right]\log \left[1+\left(\frac{v_{EP}}{v_{c}}\right)^{3}\right]}-\frac{v_{c}^{3}}{v_{c}^{3}+v^{3}}\right)\cdot 3\frac{d}{d\,\psi}\log v_{c}\Bigg\}F_{0}
    \\
    \\
    \frac{d}{d(m_{EP}\mathcal{E})}F_{0}=-\frac{3 v}{v_{c}^{3}+v^{3}}\frac{F_{0}}{m_{EP}}
    \\
    \\
    \frac{d}{d v_{\parallel}}F_{0}=0
    \end{cases}
    \label{EqC6:derivative}
\end{equation}
where $v=\sqrt{2\mathcal{E}}$ and:
\begin{equation}
    \frac{d v_{c}}{d\,\psi}=v_{c}\left[\frac{d}{d\,\psi}\log v_{th,e}-\frac{1}{3}\frac{d}{d\,\psi}\log n_{e}+\frac{1}{3}\left(\sum_{l}\frac{z_{l}^{2}}{m_{l}}\frac{d n_{l}}{d\,\psi}\right)\cdot \left(\sum_{p}\frac{z_{p}^{2}}{m_{p}n_{p}}\right)\right]\quad ,
    \label{EqC6:derivative_vc}
\end{equation}
where $\psi$ is the normalized poloidal flux.
In the notation in use:
\begin{equation}
    \mathcal{E}=\frac{v^{2}}{2}=\frac{v^{2}_{\parallel}}{2}+\mu B\quad \text{with}\quad \mu=\frac{v^{2}_{\perp}}{2 B}\quad \text{and}\quad \mathcal{E}_{EP}=m_{EP}\,\mathcal{E}\quad .
\end{equation}
We conclude this section introducing the concept of  \say{equivalent temperature} $T_{EP}$ that is the temperature of a Maxwellian distribution function $F_{M}$ for the EPs that satisfies the following requirement:
\begin{equation}
    \int\,d^{3}\Vec{v}\,v^{2}\,F_{M}=\int\,d^{3}\Vec{v}\, v^{2}\,F_{0}\quad \text{with}\quad  \int\,d^{3}\Vec{v}\, v^{2}\,F_{M}=3\,n_{EP}(r)\frac{T_{EP}(r)}{m_{EP}}\quad 
    \label{EqC6:equiv_temp0}
\end{equation}
from which it follows:
\begin{equation}
    T_{EP}(r)=\frac{2}{3}\frac{I_{4}\left(\frac{v_{c}(r)}{v_{EP}}\right)}{I_{2}\left(\frac{v_{c}(r)}{v_{EP}}\right)}\mathcal{E}_{EP}\quad .
    \label{EqC6:equiv_temp1}
\end{equation}
\subsection{Implementation in ORB5}\label{Sec:ORB5_implementation}
\begin{figure}[!h]
    \centering
    \includegraphics[width=0.8\textwidth]{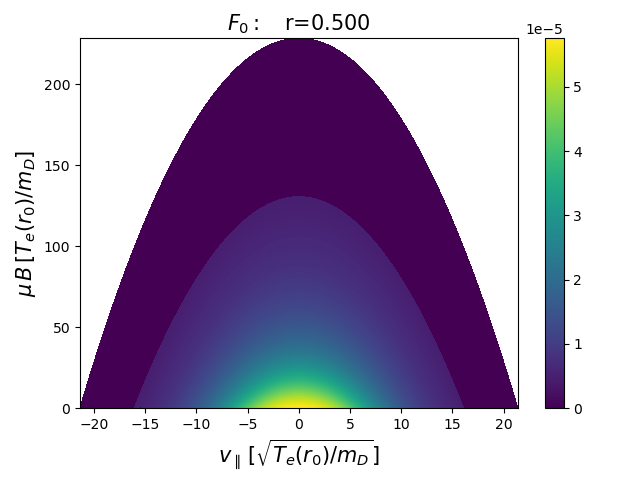}
    \caption{Slowing-down distribution implemented in ORB5 for the EPs. The dependence in velocity-space at a fixed radial position is shown.
    }
    \label{FigC6:slowing_v}
\end{figure}

\begin{figure}[!h]
    \centering
    \includegraphics[width=0.75\textwidth]{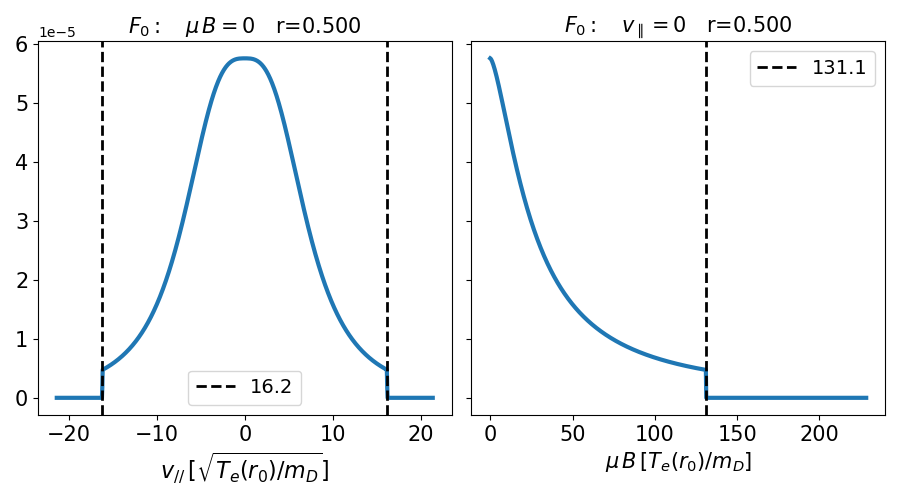}
    \caption{Slices of the equilibrium distribution function obtained from Fig.~\ref{FigC6:slowing_v}.\\ \textbf{Left:} Dependence against $v_{\parallel}$: $F_{0}(r=0.5,v_{\parallel},\mu B=0)$.  \textbf{Right:} Dependence against $\mu B$: $F_{0}(r=0.5,v_{\parallel}=0,\mu B)$.\\ The vertical black dashed lines correspond to the cuts in velocity space present because of the Heaviside theta contained in eq.~(\ref{EqC6:slowing2}).}
    \label{FigC6:slowing_cut}
\end{figure}

\begin{figure}[!h]
    \centering
    \includegraphics[width=0.75\textwidth]{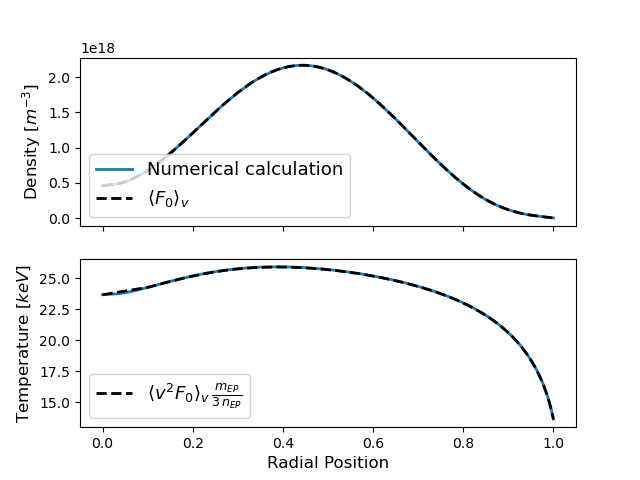}
    \caption{
    \textbf{Top:} EP Radial density profile corresponding to Fig.~\ref{FigC4:NLEDqprofile} left.  \textbf{Bottom:} EP equivalent temperature, cf. eq.~(\ref{EqC6:equiv_temp1}).\\ 
    $\langle ...\rangle_{\vec{v}}$ indicates integration in velocity space. The moments of the equilibrium distribution function $F_{0}$ implemented in ORB5 and calculated through numerical integration (blue curves) are compared with the expected values of the analytical expressions (black dashed lines).}
    \label{FigC6:moment_distribution}
\end{figure}

\begin{figure}[!h]
    \centering
    \includegraphics[width=0.75\textwidth]{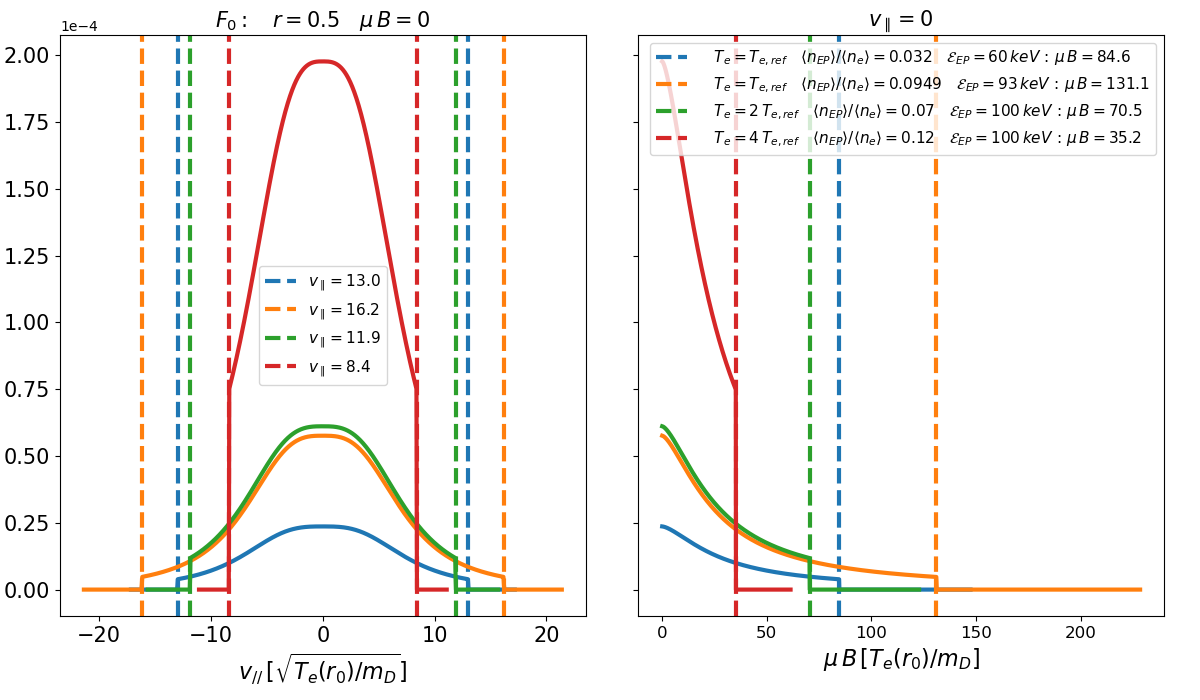}
    \caption{Slices of equilibrium distribution functions obtained considering different EP concentration, values of electron temperature and injection energy. $T_{e,ref}$ indicates the reference electron temperature of the NLED-AUG case (cf. Fig.~\ref{FigC4:NLEDplasmaprofiles} left).\\ \textbf{Left:} Dependence against $v_{\parallel}$. \textbf{Right:} Dependence against $\mu B$.}
    \label{Fig:slowing_cut}
\end{figure}

\begin{figure}[!h]
    \centering
    \includegraphics[width=0.75\textwidth]{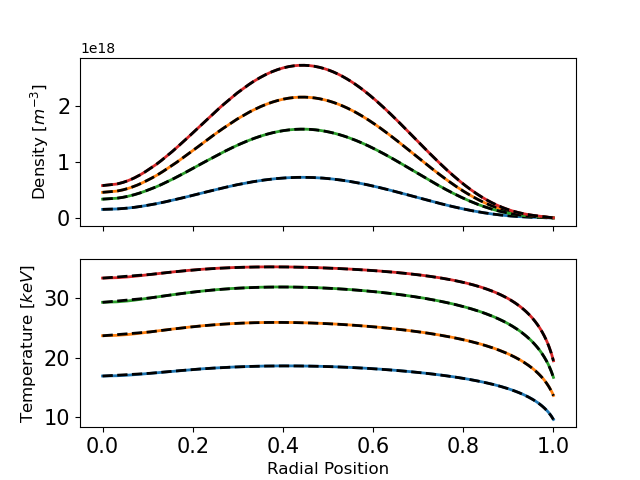}
    \caption{
    \textbf{Top:} EP Radial density profiles. \textbf{Bottom:} EP equivalent temperature profiles.\\ 
    The black-dashed lines correspond to the analytical expressions, while the coloured lines are calculated through numerical integration of the implemented distribution functions. The coloured lines are obtained with the plasma parameters  considered to compute the slices of the same colour in Fig.~\ref{Fig:slowing_cut}}
    \label{Fig:moment_distribution}
\end{figure}

As it has been already anticipated in Sec.~\ref{Sec:Num_model}, in ORB5 all the quantities are normalized with respect to reference parameters. In particular, velocities are given in units of the sound velocity $c_{s}$:
\begin{equation}
    c_{s}=\sqrt{\frac{T_{e}(r_{0})}{m_{D}}}
\end{equation}
defined as the square root of the electron temperature considered at a chosen reference position $r_{0}$, divided by the mass of the main ion species that, for the NLED-AUG case, is constituted by deuterium ions. Here the reference radial position is on-axis $r_{0}=0$ and $T_{e}(r_{0})=0.7088\,$keV . In Fig.~\ref{FigC6:slowing_v} the velocity dependence in $(v_{\parallel},\mu\,B)$-space of the slowing-down distribution function at $r=0.5$ is shown. This has been taken from an ORB5-simulation. It has been obtained considering the bulk plasma parameters of the NLED-AUG case (see Sec.~\ref{Sec:NLED_scenario}) as well as the reference EPs parameters: off-axis radial density profile in Fig.~\ref{FigC4:NLEDqprofile} left having a concentration of $\langle n_{EP}\rangle/\langle n_{e}\rangle=0.0949$, injected with injection energy $\mathcal{E}_{EP}=93\,$keV.
\\
In Fig.~\ref{FigC6:slowing_cut} slices obtained from Fig.~\ref{FigC6:slowing_v} are shown. There we can observe the cuts in velocity-space introduced  because of the presence of the Heaviside function in \cref{EqC6:slowing2}:
\begin{equation}
    v_{\parallel}\approx 16.2\,c_{s}\quad ,\quad \mu\,B \approx 131.1\, c_{s}^{2}\quad .
\end{equation}
Using the appropriate conversion factor the interested reader can easily verify that these cuts correspond to an injection energy of $\mathcal{E}_{EP}=93\,$keV. In Fig.~\ref{FigC6:moment_distribution} we show the EP radial density profile (on the top, corresponding to Fig.~\ref{FigC4:NLEDqprofile} left) and the equivalent temperature (on the bottom, cf. \cref{EqC6:equiv_temp1}). The analytical expressions (black dashed lines) are compared with the corresponding moments (blue lines) calculated through numerical integration of the distribution function implemented in ORB5. The observed good agreement between the analytical expressions and the numerical calculations is a further proof of the correct implementation of the new non-Maxwellian background distribution function in the code.
\\
In Fig.~\ref{Fig:slowing_cut} slices of slowing-down distribution functions obtained considering different EP concentration, electron temperature and injection energy are shown. The corresponding moments are shown in Fig.~\ref{Fig:moment_distribution}. Also there a good agreement is observed between the analytical expressions (black dashed lines) and the numerical integrals of the implemented slowing-down.

\section{Comparison with the experiments: application to the NLED-AUG case}\label{Sec:NLED}
\begin{figure}[!h]
    \centering
    \includegraphics[width=0.8\textwidth]{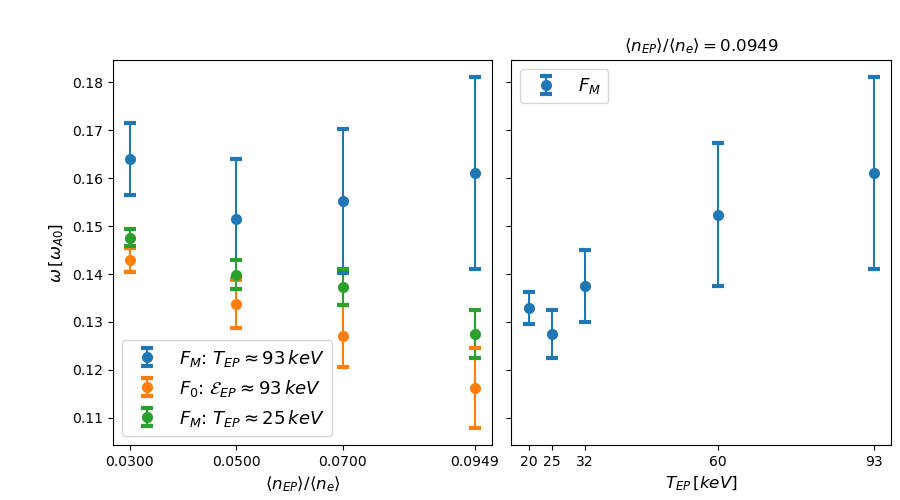}
    \caption{Real frequencies determined in the growing exponential phase of the dominant modes in  drift-kinetic simulations (no FLR effects considered). \textbf{Left:} Scan against the EP concentration for Maxwellian ($F_{M}$) and slowing-down ($F_{0}$) distribution functions.\\ \textbf{Right:} Scan against the EP temperature in simulations where the EPs have Maxwellian distribution functions.}
    \label{FigC6:scan_omega_dk}
\end{figure}

\begin{figure}[!h]
    \centering
    \includegraphics[width=1.0\textwidth]{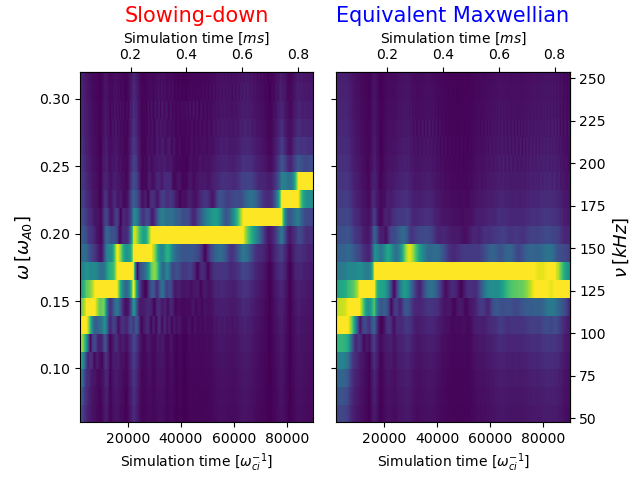}
    \caption{Spectrograms calculated from ORB5 simulations with the EPs modelled with different distribution functions (as indicated in the plot titles). The Fourier transform of the signal has been calculated in the temporal range $t\in [t_{0},t_{0}+\Delta t]$ with $\Delta t=10^{4}\,\omega_{ci}^{-1}$. The low boundary $t_{0}$ has been varied continuously as indicated in the $x$-axis of the plot.}
    \label{FigC6:spectrogram}
\end{figure}

We discuss here the mode dynamics observed in numerical simulations where the NLED-AUG scenario (see Sec.~\ref{Sec:NLED_scenario}) has been taken into account. In Tab.~\ref{TabC6:NLED_sims} the main parameters considered in the simulations are reported. Their choice is motivated by convergence studies (see Appendix \ref{AppendixSlowing}). As it is indicated in Tab.~\ref{TabC6:NLED_sims}, we will investigate here only  the dynamics of modes with toroidal mode number $n=1$. This because, as anticipated in Sec.~\ref{Sec:NLED_scenario}, from experimental measurements this is known to be   the most unstable AM. The dynamics of zonal structures ($n=0$) is then not treated in the present work.
\begin{table}[!h]
    \centering
    \begin{tabular}{||c|c|c|c|c|c|c||}
        \hline
         $\Delta t\,[\omega_{ci}^{-1}]$& $nptot_{D,e,EP}\cdot 10^{7}$ & $N_{r}$ & $N_{\theta^{*}}$ & $N_{\varphi}$ & $nfilt$ & $\Delta m$ \\
         \hline
         $4$ & $3,12,3$ & $3000$ & $64$ & $32$ & $1$ & $7$ \\
         \hline
    \end{tabular}
    \caption{Main simulation parameters: time step ($\Delta t$), number of markers for particle species ($nptot$). Radial/poloidal/toroidal grid points $(N_{r},N_{\theta^{*}},N_{\varphi})$, toroidal ($nfilt$) and poloidal width ($\Delta m$) of the field aligned Fourier filter (see Sec.~\ref{Sec:Num_model}).}
    \label{TabC6:NLED_sims}
\end{table} 
\\
We begin studying the mode dynamics observed in the exponential growth phase of the dominant mode and compare the results obtained with a Maxwellian and a slowing-down distribution. For the moment we run ORB5 neglecting the FLR effects of all the particle species, i.e. considering a drift-kinetic model. In all the simulations analyzed the dominant mode, in the exponential growth phase, is an Alfv\'en mode with scalar potential dominated by its components $(m,n)=(2,1)$ and peaked around $r=0.2$. 
\begin{figure}[!h]
    \centering
    \includegraphics[width=0.8\textwidth]{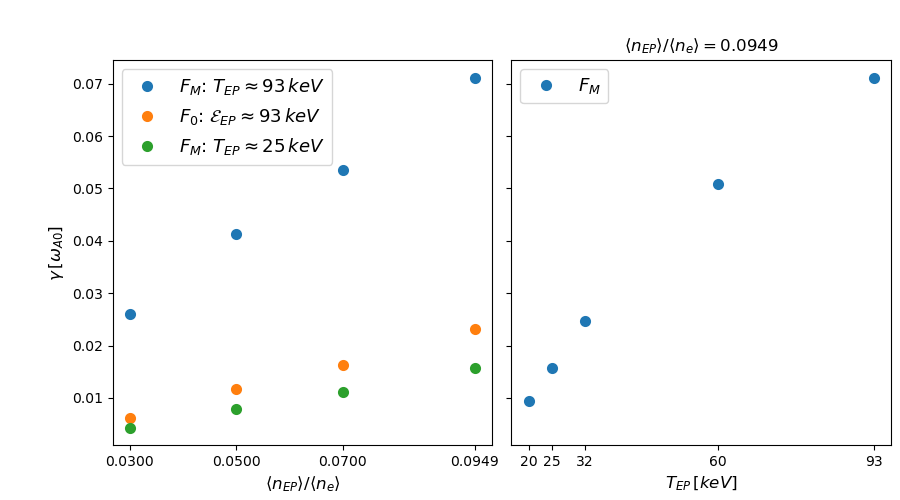}
    \caption{Growth rates determined in drift-kinetic simulations (no FLR effects). \textbf{Left:} Scan against the EP concentration for Maxwellian ($F_{M}$) and slowing-down ($F_{0}$) distribution functions. $T_{EP}=25\,$keV represents the value on the center of the equivalent temperature (cf. \cref{EqC6:equiv_temp1}) corresponding to an injection energy of $\mathcal{E}_{EP}=93\,$keV. \textbf{Right:} Scan against the EP temperature in simulations where the EPs have been modelled through Maxwellian distribution functions.}
    \label{FigC6:scan_gamma_dk}
\end{figure}
\\
In Fig.~\ref{FigC6:scan_gamma_dk} left the growth rates determined in a scan against the EP concentration are shown. There we compare the growth rates obtained in simulations where the EPs have been modelled with Maxwellian distribution functions ($F_{M}$) with those obtained modelling the EPs with the slowing-down distribution function ($F_{0}$). In Fig.~\ref{FigC6:scan_gamma_dk} right the growth rate dependence against the EPs temperature (with constant radial profile) is shown in a scan where the EPs have been modelled via Maxwellian distribution functions. In Fig.~\ref{FigC6:scan_omega_dk} we display  the real frequencies of the dominant modes observed in the corresponding simulations that have produced the growth rates reported in Fig.~\ref{FigC6:scan_gamma_dk}. Only in Fig.~\ref{FigC6:scan_omega_dk} the error bars are present, since the errors committed determining  the growth rates in Fig.~\ref{FigC6:scan_gamma_dk} are negligible.
\\
In the scans presented in Fig.~\ref{FigC6:scan_gamma_dk} and Fig.~\ref{FigC6:scan_omega_dk}, the case closer to the experimental conditions is that where the EPs have been modelled with slowing-down distribution function and concentration of $\langle n_{EP}\rangle/\langle n_{e}\rangle=0.0949$. Its values of growth rate and frequency are reported in  Tab.~\ref{TabC6:NLED_sims_dk}. There, they are compared with those obtained in a simulation where the EPs have been modelled with a Maxwellian distribution function with $T_{EP}=25\,$keV. This  corresponds to the value of the equivalent temperature around $r\approx 0.5$ (cf. with Fig.~\ref{FigC6:moment_distribution}).
\begin{table}[!h]
    \centering
    \begin{tabular}{||c|c|c||}
        \hline
         Distribution function & $\gamma\,[\omega_{A0}]$ & $\omega\,[\omega_{A0}]$  \\
         \hline
         Slowing-down, $\mathcal{E}_{EP}=93\,$keV & $0.023$ & $0.116\pm 0.008$  \\
         Maxwellian, $T_{EP}=25\,$keV & $0.016$ & $0.128\pm 0.005$  \\
         \hline
    \end{tabular}
    \caption{Growth rates $\gamma$ and frequencies $\omega$ of the dominant modes, determined in drift-kinetic simulations (no FLR effects included). The reference EP concentration of $\langle n_{EP}\rangle/\langle n_{e}\rangle=0.0949$ is here considered.}
    \label{TabC6:NLED_sims_dk}
\end{table} 
\\
Referring to Tab.~\ref{TabC6:NLED_sims_dk}, we can then conclude that the choice of the slowing-down distribution function over a Maxwellian distribution function with temperature close to the equivalent temperature, does not affect significantly the frequency of driven AM, although quantitative differences as large as almost $40\,\%$ are found in the growth rate.
\\
\yFigTwo{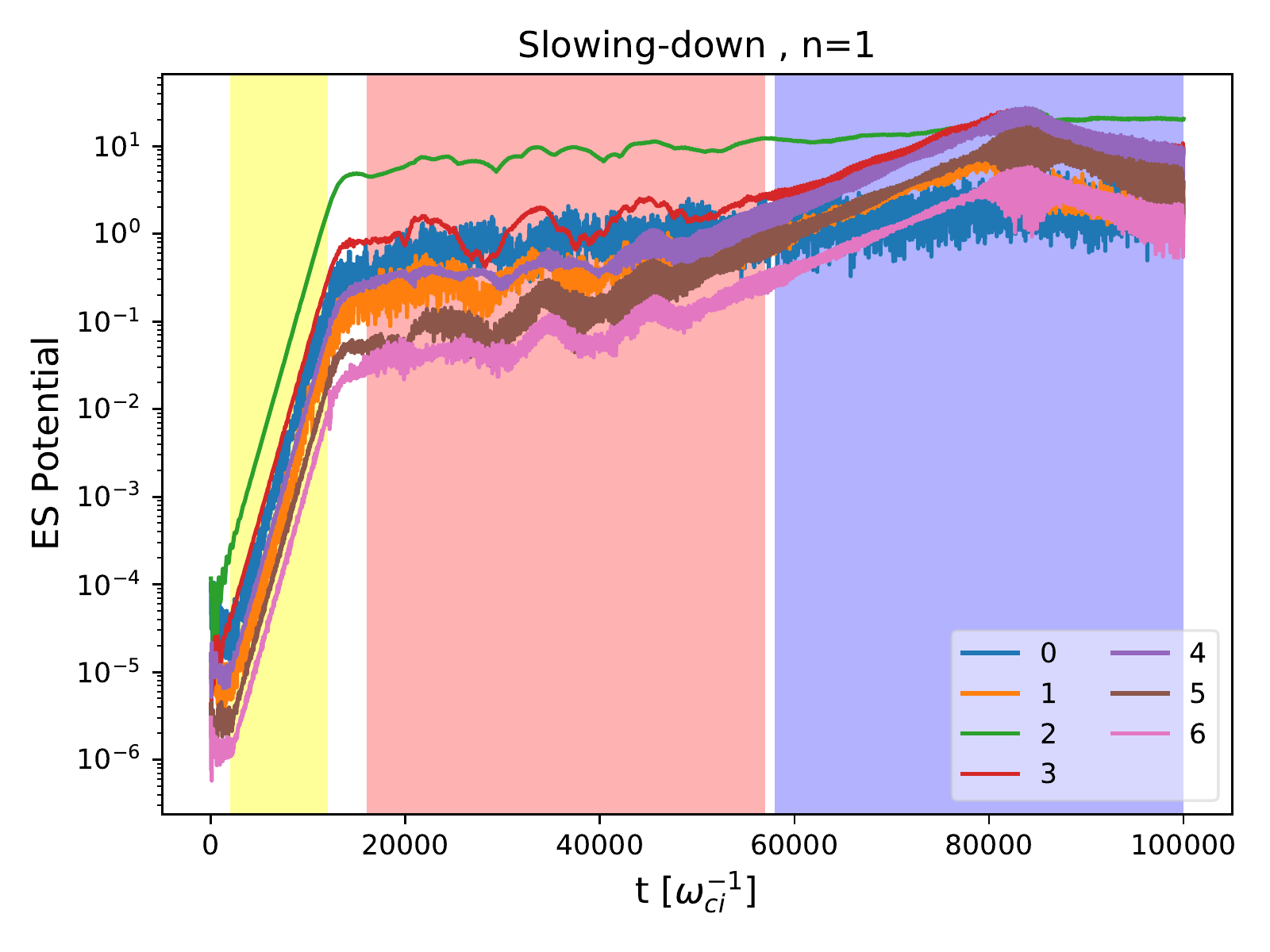}{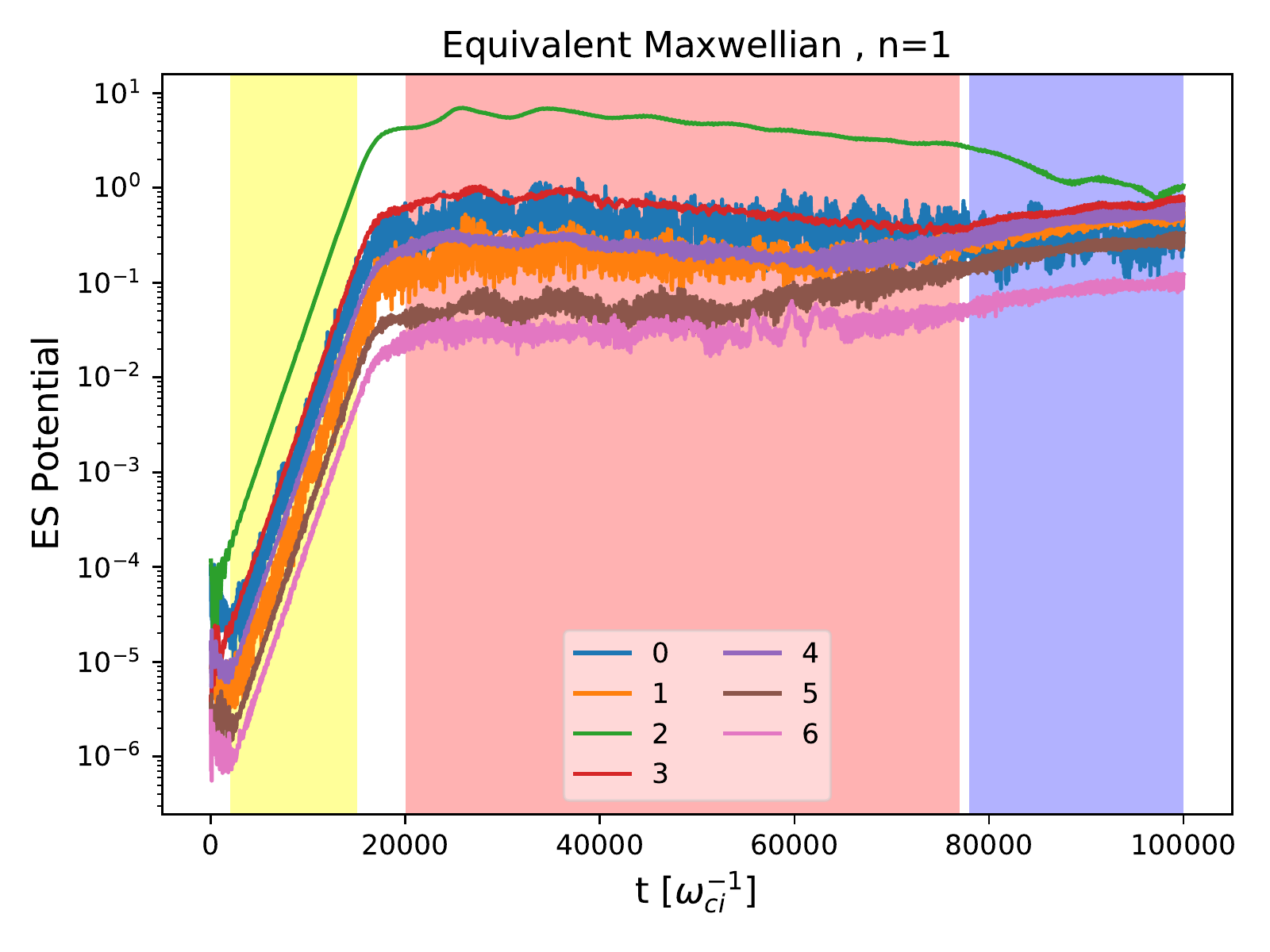}{Temporal evolution of the dominant poloidal harmonics ($m$-numbers indicated in the legends) in the simulations. \textbf{Left}: EPs modelled with slowing-down distribution and $\mathcal{E}_{EP}=93\,$keV. \textbf{Right:} EPs modelled with the equivalent Maxwellian.}{Fig:dynamics_diff}
\\
We describe now in detail the mode dynamics of the AM observed in simulations where FLR effects are included for bulk ions and EPs (the model is gyrokinetic) and the reference EP concentration of $\langle n_{EP}\rangle/\langle n_{e}\rangle=0.0949$ is taken into account. Here we will not just limit ourselves to the investigation of the exponential growth phase of the AM, but also to its nonlinear phase. There the AM mode structure evolves radially, presenting peaks close to the edge. Therefore we have increased the radial resolution to $N_{r}=6000$ in order to correctly resolve those peaks. Simultaneously since in ORB5 the numerical noise is proportional to the number of radial grid points $N_{r}$ \cite{bottino_sonnendrucker_2015}, we have increased the number of markers to $nptot_{E,e,EP}=(18,72,18)\cdot 10^{7}$.
\\
The temporal dynamics observed in a simulation where the EPs have been modelled with the slowing-down distribution function with $\mathcal{E}_{EP}=93\,$keV, is compared  with that observed in a simulation where the EPs are modelled with the equivalent Maxwellian (the equivalent temperature in  Fig.~\ref{FigC6:moment_distribution} is taken now into account). The former is shown in Fig.~\ref{Fig:dynamics_diff} left, while the latter is shown in Fig.~\ref{Fig:dynamics_diff} right. In the two plots in Fig.~\ref{Fig:dynamics_diff} we have highlighted the temporal intervals corresponding to: the exponential growth phase (yellow), the early nonlinear phase (pink) and the deep nonlinear phase (blue). Note that both the simulations cover the same temporal range: $t\in [0;100000]\,\omega_{ci}^{-1}$.
\begin{table}[!h]
    \centering
    \begin{tabular}{||c|c|c||}
        \hline
         Distribution function & $\gamma\,[\omega_{A0}]$ & $\omega\,[\omega_{A0}]$  \\
         \hline
         Slowing-down, $\mathcal{E}_{EP}=93\,$keV & $0.019$ & $0.133\pm 0.007$  \\
         Equivalent Maxwellian & $0.015$ & $0.138\pm 0.005$  \\
         \hline
    \end{tabular}
    \caption{Growth rates $\gamma$ and frequencies $\omega$ of the dominant modes, determined in gyrokinetic simulations (FLR effects of the EPs are retained). The reference EP concentration of $\langle n_{EP}\rangle/\langle n_{e}\rangle=0.0949$ is here considered.}
    \label{TabC6:NLED_sims_gyr}
\end{table} 
\\
The growth rate and frequencies determined in the exponential growth phase of the two simulations in Fig.~\ref{Fig:dynamics_diff} are reported in Tab.~\ref{TabC6:NLED_sims_gyr}, while the corresponding mode structures and frequency spectra are shown in  Fig.~\ref{Fig:dynamics_diff_linear}. The dominant AM here, as in the drift-kinetic simulations, is an EPM whose scalar potential is dominated by the Fourier component $(m,n)=(2,1)$ peaked around the radial position $r\approx 0.2$. Its frequency lies almost on the continuum spectrum calculated with the linear gyrokinetic code LIGKA \cite{LIGKA} (see red dashed lines in Fig.~\ref{Fig:dynamics_diff_linear} on the bottom). We can conclude also here that, as for the drift-kinetic simulations, the choice of the equivalent Maxwellian over the slowing-down distribution function, does not modify significantly the kind of driven AM investigated (mode structure and frequency). However, referring to Tab.~\ref{TabC6:NLED_sims_gyr}, we still observe a growth rate quantitative difference that in this case is as large as almost $30\,\%$ of the value of the growth rate obtained in the simulation with the EPs modelled with the equivalent Maxwellian.
\yFigFour{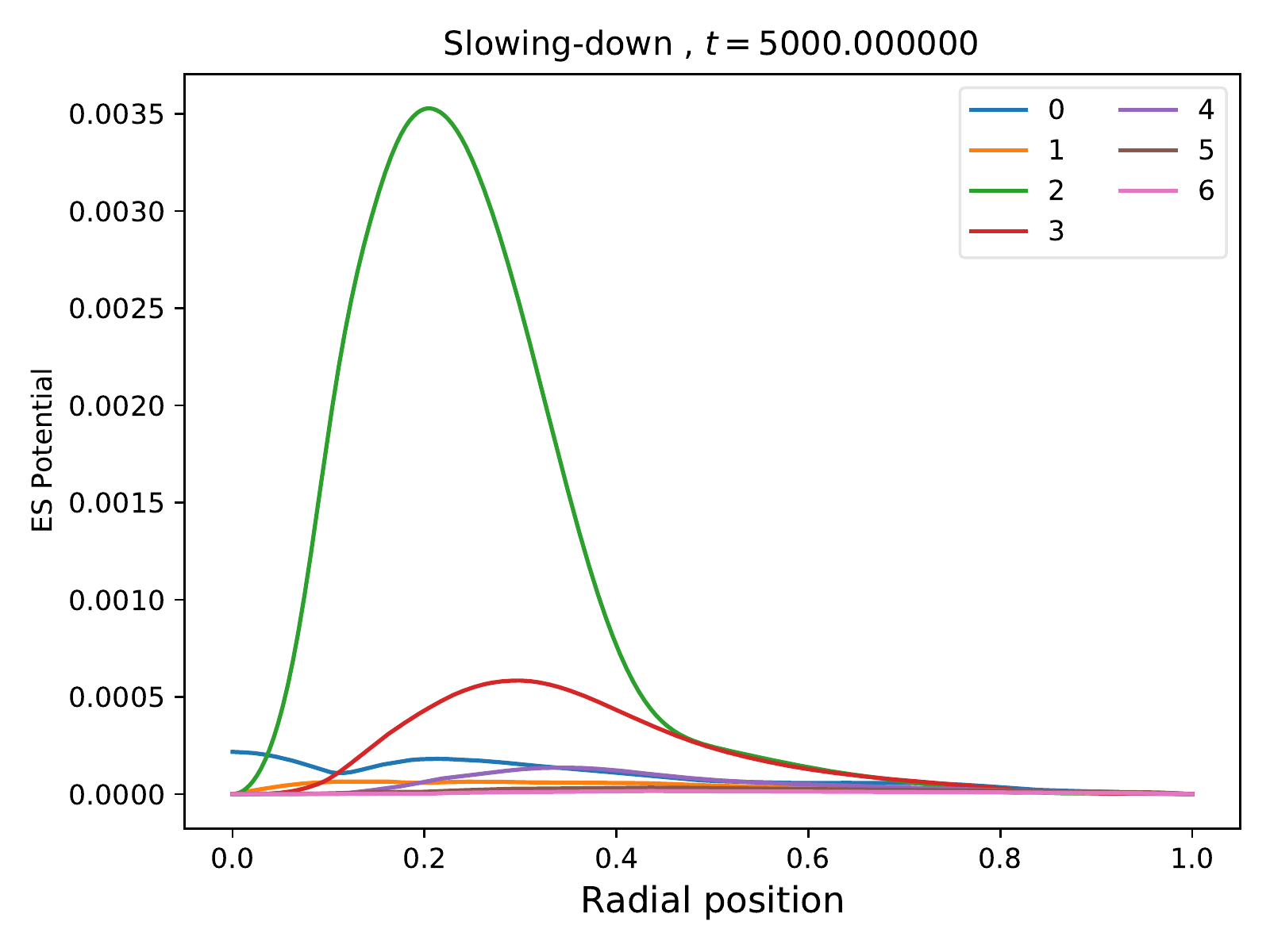}{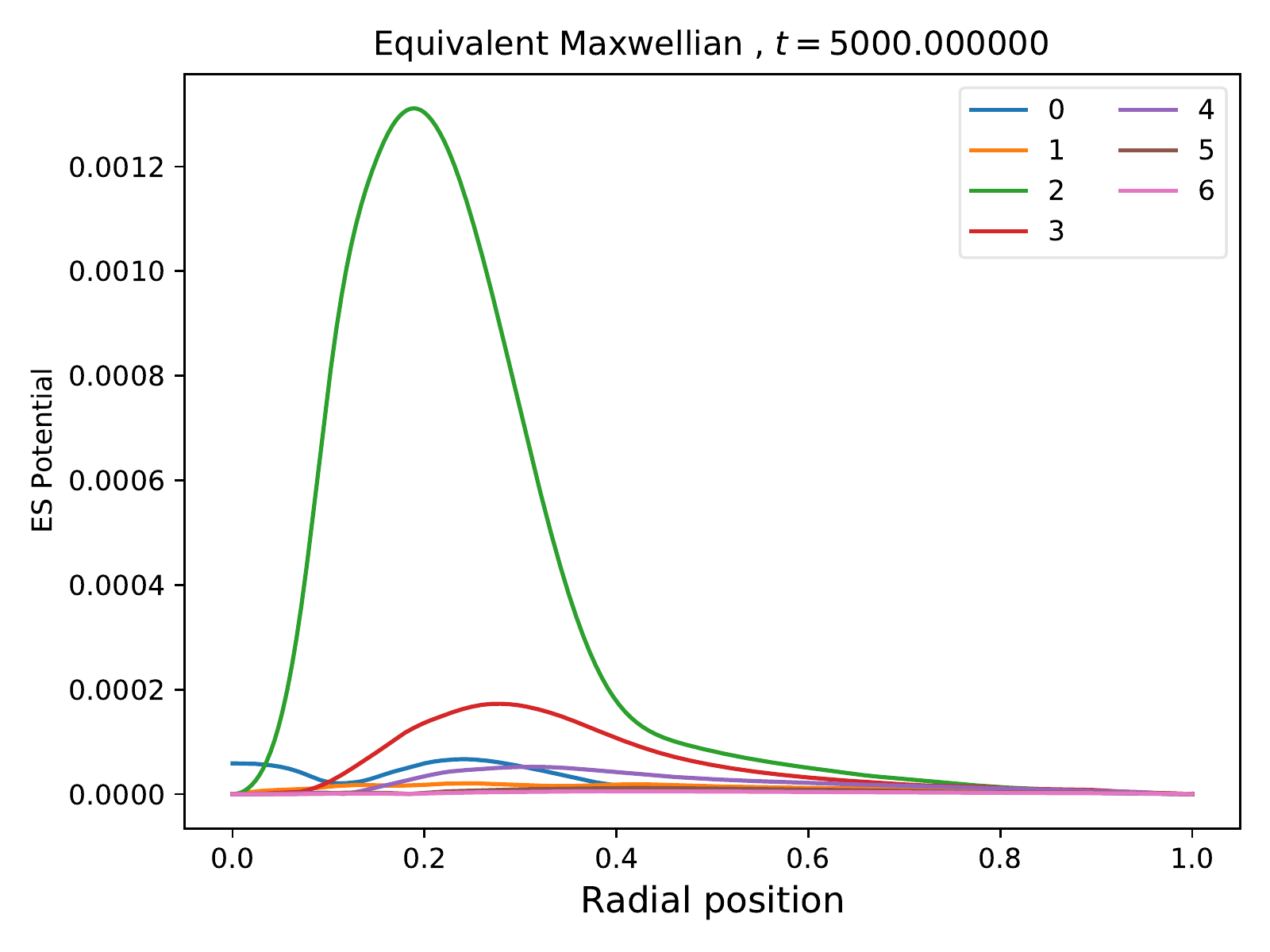}{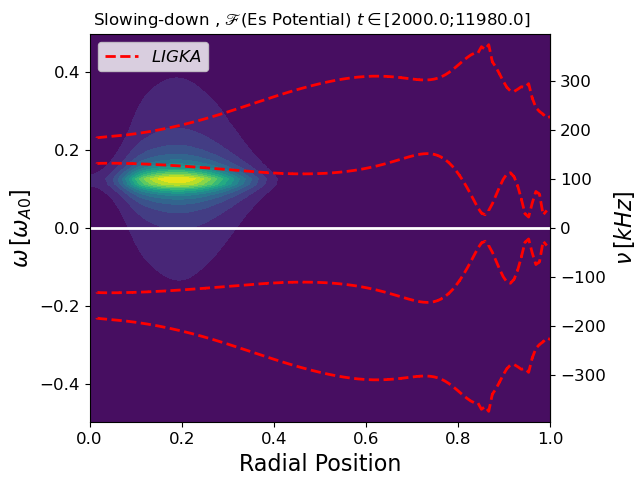}{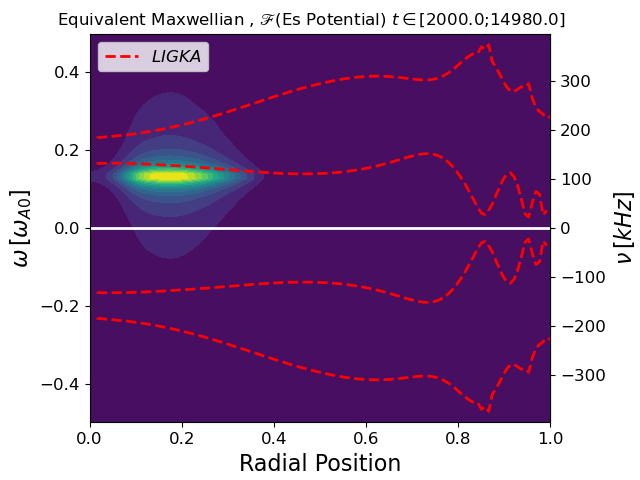}{Mode structures and frequency spectra observed in the exponential growth phases (yellow regions) of the simulations in Fig.~\ref{Fig:dynamics_diff}, where the EPs have been modelled with slowing-down (\textbf{left}) and equivalent Maxwellian (\textbf{right}).}{Fig:dynamics_diff_linear}
\yFigTwo{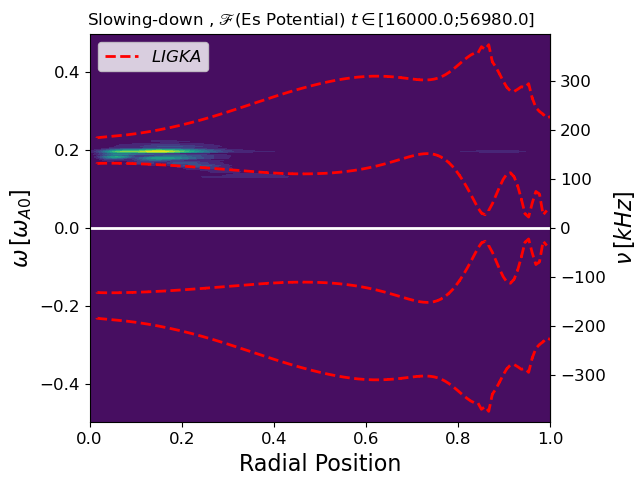}{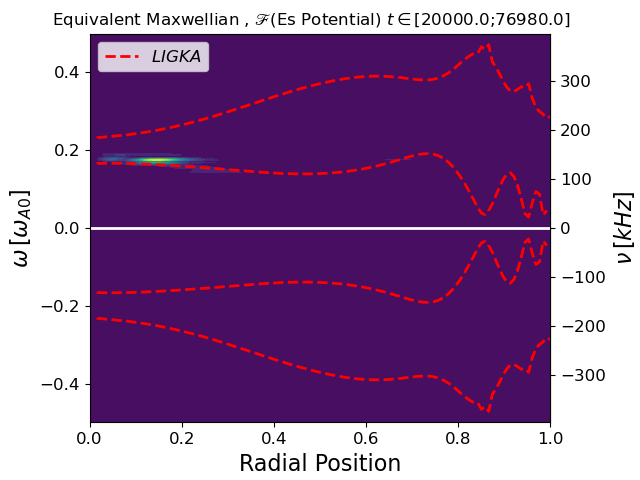}{Frequency spectra calculated in the early nonlinear phases (pink regions) of the simulations in Fig.~\ref{Fig:dynamics_diff}. \textbf{Left:} EPs modelled with slowing-down distribution function. \textbf{Right:} EPs modelled via equivalent Maxwellian.}{Fig:dynamics_nonlinear1}
\\
In Fig.~\ref{Fig:dynamics_nonlinear1} we compare the frequency spectra calculated in the early nonlinear phases (pink regions) of the two different simulations shown in Fig.~\ref{Fig:dynamics_diff}. In this temporal phase the scalar potential Fourier component $(m,n)=(2,1)$ is still the dominant one, since it is almost an order of magnitude higher than the other Fourier components. Comparing Fig.~\ref{Fig:dynamics_nonlinear1} with Fig.~\ref{Fig:dynamics_diff_linear}, we observe that both simulations catch correctly the nonlinear modification of the AM frequency passing from the linear to the nonlinear phase. In this time interval, the dominant AM frequency exhibits tiny differences modelling the EPs with slowing-down or equivalent Maxwellian. Additionally in both simulations, at the end of this phase, the Fourier component $(m,n)=(3,1)$ begins to grow, even though with different growth rate, as indicated in Tab.~\ref{Tab:nonlinear_3_growth}.
\begin{table}[!h]
    \centering
    \begin{tabular}{||c|c||}
        \hline
         Distribution function & $\gamma\,[\omega_{A0}]$ \\
         \hline
         Slowing-down, $\mathcal{E}_{EP}=93\,$keV & $0.002$  \\
         Equivalent Maxwellian & $0.00053$   \\
         \hline
    \end{tabular}
    \caption{Growth rates $\gamma$ of $(m,n)=(3,1)$ in the deep nonlinear phase (blue regions in Fig.~\ref{Fig:dynamics_diff}). Gyrokinetic simulations (FLR effects of the EPs are retained). The reference EP concentration of $\langle n_{EP}\rangle/\langle n_{e}\rangle=0.0949$ is here considered.}
    \label{Tab:nonlinear_3_growth}
\end{table} 
\\
\yFigTwo{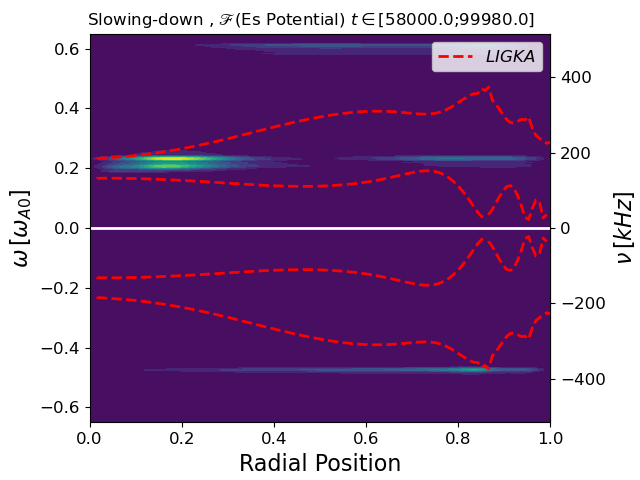}{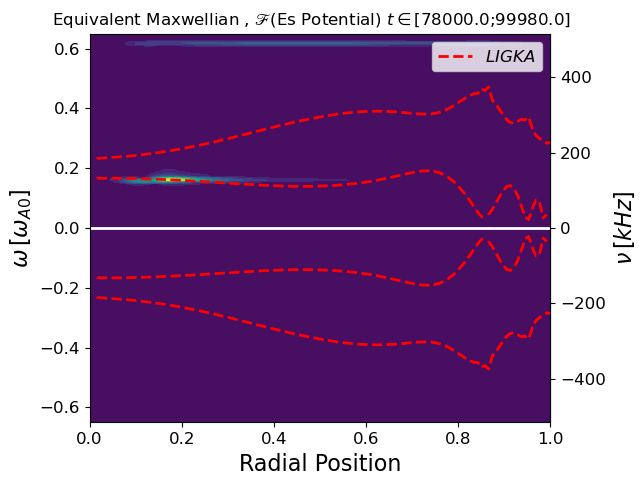}{Frequency spectra calculated in the deep nonlinear phases (blue regions) of the simulations in Fig.~\ref{Fig:dynamics_diff}. \textbf{Left:} EPs modelled with slowing-down distribution function. \textbf{Right:} EPs modelled via equivalent Maxwellian.}{Fig:dynamics_nonlinear2}
\\
In Fig.~\ref{Fig:dynamics_nonlinear2} the frequency spectra calculated in the deep nonlinear phases (blue regions in Fig.~\ref{Fig:dynamics_diff}) are shown. In these temporal phases the scalar potential Fourier component $(m,n)=(3,1)$ reaches an amplitude comparable to that of the dominant $(m,n)=(2,1)$.  Passing from  Fig.~\ref{Fig:dynamics_nonlinear1} to Fig.~\ref{Fig:dynamics_nonlinear2} we observe that in the simulation where the EPs have been modelled with the slowing-down (left), a higher frequency modification of the AM is detected with respect to the right plots where the EPs had been modelled with the equivalent Maxwellian. In fact in Fig.~\ref{Fig:dynamics_nonlinear2} left the AM oscillates mainly at the TAE frequency. We remind here that this is an AE whose frequency lies inside the gap created by the branches of the continuum $(m_{0},n)$ and $(m_{0}+1,n)$ at the radial position $r_{0}$ that labels the rational surface satisfying $q(r_{0})=(2\,m_{0}+1)/(2\,n)$ \cite{CHENG198521} where two close poloidal harmonics interact. For the case under investigation: $m_{0}=2$, $n=1$ and $r_{0}=0.738$ (cf. Fig.~\ref{FigC4:NLEDqprofile} right). We emphasize that in Fig.~\ref{Fig:dynamics_nonlinear2} left the AM oscillates almost at every radial position at the TAE frequency while, in Fig.~\ref{Fig:dynamics_nonlinear2} right, such a higher  frequency modification is not observed. 
Finally we note in  Fig.~\ref{Fig:dynamics_nonlinear2} the presence, in this deep nonlinear phase, of higher frequencies in the EAE gap: $\omega\sim 0.6,-0.5\,\omega_{A0}$. We do not further investigate the emerging of these higher frequencies as the focus of the present paper is on the study of the TAE-EPM frequency modification observed in the experiment (cf. Fig.~\ref{Fig:experimental_spectrogram}), that correspond to the dominant frequencies also in the presented simulations.
\\
We summarize the results of the AM frequency modification in  Fig.~\ref{FigC6:spectrogram}. There the spectrogram calculated in a simulation with EPs modelled with the slowing down (left) is compared with that obtained in the simulation where the EPs have been modelled with the equivalent Maxwellian (right). The spectrograms presented have been obtained through the fast Fourier transform of the scalar potential at $r=0.2$. This corresponds to the radial positions where the mode structure of the AM has a maximum. The time interval where the Fourier transform has been calculated $[t_{0};t_{0}+\Delta t]$ has been obtained varying continuously $t_{0}$ (as indicated in the $x$-axis in Fig.~\ref{FigC6:spectrogram}) and choosing $\Delta t = 10^{4}\,\omega_{ci}^{-1}$. In Fig.~\ref{FigC6:spectrogram} the time units are provided also in milliseconds, while the frequency units are provided also in kHz, to allow a better comparison with the experimental spectrogram in Fig.~\ref{Fig:experimental_spectrogram}. We notice, at first, that the total time width of the simulations is of $\approx\,1\,$ms which lies well inside the width of the TAE-EPM burst at $t=0.84\,$s (cf. Fig.~\ref{Fig:experimental_spectrogram}). Finally, the choice of the slowing-down over the equivalent Maxwellian allows a better comparison with the experiment. Through this, in fact, a broader modification of the AM frequency in the nonlinear phase is observed $100\,$kHz$ \leq\nu\leq 180\,$kHz, resulting in a better agreement with the experiment (cf. Fig.~\ref{Fig:experimental_spectrogram}).
\section{Conclusion}\label{Sec:conclusion}
In this work the study of the AM instabilities driven by EPs has been conducted for the NLED-AUG scenario. This is an interesting case obtained in the Tokamak ASDEX Upgrade where the EPs have been injected  through a NB with energy $\mathcal{E}_{EP}\approx 93\,$keV. The peculiarity of this scenario relies on the fact that it has been obtained through plasma ratios ($\beta_{EP}/\beta_{Bulk}$ and $\mathcal{E}_{EP}/T_{Bulk}$) close to those that are going to be met in future fusion machines.\\
Studies of the AM dynamics in the NLED-AUG case have already been conducted through numerical simulations \cite{VANNINI2020,VANNINI2021,Vlad2021}. The main novelty in the present work is represented by the fact that the EPs have been modelled thorough an equilibrium isotropic slowing-down distribution function.
Through it, we have been able to go closer to the experimental conditions. In particular we have obtained a good quantitative agreement with the experimental spectrogram doing a further step with respect to previous published works \cite{VANNINI2020,VANNINI2021,Vlad2021}. There, in fact, a good qualitative agreement with the experiment has been obtained and the EPs have been modelled with Maxwellian or double-bump-on-tails, whose choice was motivated by the kind of physics we were interested to investigate. In the present work we have proved that the isotropic slowing-down distribution function has been correctly implemented in ORB5. We have also shown that its choice allows a more accurate description of  the frequency modification of the AM in the nonlinear phase, while the choice of an equivalent Maxwellian underestimate it. The key conclusion is that with the inclusion of the FLR effects we have been able to achieve a good quantitative agreement with the experimental spectrogram. These results are already encouraging and prove that we are able to reproduce the relevant physics behind the experiments in an attempt to become predictive about future fusion scenarios and pave the path to new studies that have to be carried out to catch more of the nonlinear aspects in the NLED-AUG case.\\
In future works we will retain the nonlinearities in all the particle species and we will consider the interaction between the AM and the EGAM, to describe the triggering of the EGAM by the TAE-EPM burst. An attempt on this has already been done in Ref.~\cite{VANNINI2021} where the EPs have been modelled with a double-bump-on-tail. Moreover the AM and EGAM interaction will be studied considering a $\xi$-dependent slowing-down distribution function  \cite{Rettino} in order to retain the anisotropy in velocity space needed to drive the EGAM.

\section{Acknowledgments}
Simulations presented in this work were performed on the CINECA Marconi supercomputer within the ORBFAST, OrbZONE and TSVV10 projects.
The authors acknowledge stimulating discussions with Z. Lu, F. Zonca, S. Briguglio, I. Novikau and  A. Di Siena. This work was partly performed in the frame of the “Multi-scale Energetic Particle Transport in Fusion Devices” ER project.
This work has been carried out within the framework of the EUROfusion Consortium, funded by the European Union via the Euratom Research and Training Programme (Grant Agreement No 101052200 — EUROfusion). Views and opinions expressed are however those of the author(s) only and do not necessarily reflect those of the European Union or the European Commission. Neither the European Union nor the European Commission can be held responsible for them.

\clearpage
\appendix
\section{Slowing-down validation and convergence studies}\label{AppendixSlowing}
We present here a series of convergence studies used to find the reference simulation parameters in Tab.~\ref{TabC6:NLED_sims}. In Fig.~\ref{FigC6:scan_dt}, Fig.~\ref{FigC6:scan_nptot}, Fig.~\ref{FigC6:scan_nptote} and Fig.~\ref{FigC6:scan_Ns} we show the discrepancies $\sigma$ of the determined quantities $X$ from the most accurate value, that is:
\begin{equation}
    \sigma\,[\%]=\frac{X-X_{most\,accurate\,vale}}{X_{most\,accurate\,vale}}\cdot 100\quad .
    \label{EqAppD}
\end{equation}
The most accurate value $X_{most\,accurate\,value}$ corresponds to the value (almost) at convergence. Its discrepancy, because of the definition in eq.~(\ref{EqAppD}), is $\sigma=0\,\%$. In Fig.~\ref{FigC6:scan_dt}, Fig.~\ref{FigC6:scan_nptot}, Fig.~\ref{FigC6:scan_nptote} and Fig.~\ref{FigC6:scan_Ns} the orange points have been obtained in simulations where the reference parameters in Tab.~\ref{TabC6:ref_sims} have been considered.
\begin{table}[!h]
    \centering
    \begin{tabular}{||c|c|c|c|c|c|c||}
        \hline
         $\Delta t\,[\omega_{ci}^{-1}]$& $nptot_{D,e,EP}\cdot 10^{7}$ & $N_{r}$ \\
         \hline
         $4$ & $3,12,3$ & $1000$\\
         \hline
    \end{tabular}
    \caption{Reference simulation parameters: time step ($\Delta t$), number of markers for particle species ($nptot$) and  radial grid points $(N_{r})$, cf. Sec.~\ref{Sec:Num_model}.}
    \label{TabC6:ref_sims}
\end{table} 
\\
In all the simulations here presented FLR effects are retained and the reference EP concentration of $\langle n_{EP}\rangle/\langle n_{e}\rangle=0.0949$ is considered. 
The quantities $X$ here studied (growth rates $\gamma$ and frequencies $\omega$) have been determined in the exponential growth phase of the simulations. The scans here presented are against: the time width $\Delta t$, the number of markers $nptot$ and the number of radial grid points $N_{r}$ (cf. Sec.~\ref{Sec:Num_model}). The growth rate and frequency obtained using the reference parameters in Tab.~\ref{TabC6:ref_sims} correspond to:
\begin{equation}
    \gamma_{ref} = 0.0197\,\omega_{A0} \quad ,\quad \omega_{ref}=0.133\,\omega_{A0}\quad .
\end{equation}

\yFigTwo{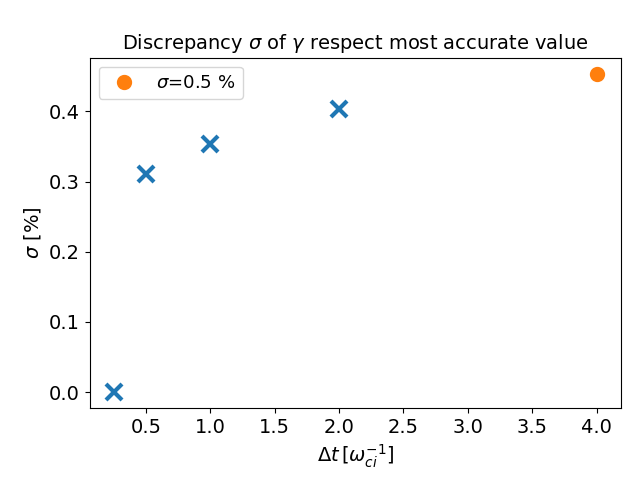}{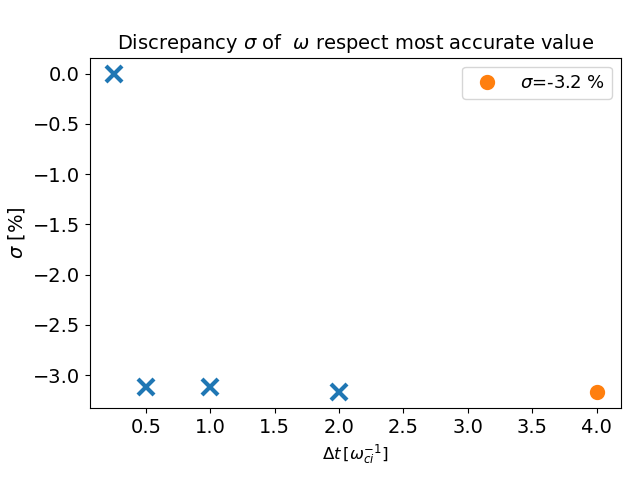}{Simulations with $\#$ of markers$_{D,e,EP}=(3,12,3)\cdot 10^{7}$ and $N_{r}=1000$. Scan against the time step $\Delta t$ in use in the simulations. In the plots, the discrepancies $\sigma$ of growth rate (\textbf{left}) and frequencies (\textbf{right}) with respect to the best resolved case (at $\Delta t=0.25\,\omega_{ci}^{-1}$) are shown. 
 The orange points correspond to the results obtained with the parameters shown in Tab.~\ref{TabC6:ref_sims}.}{FigC6:scan_dt}
 
 \yFigTwo{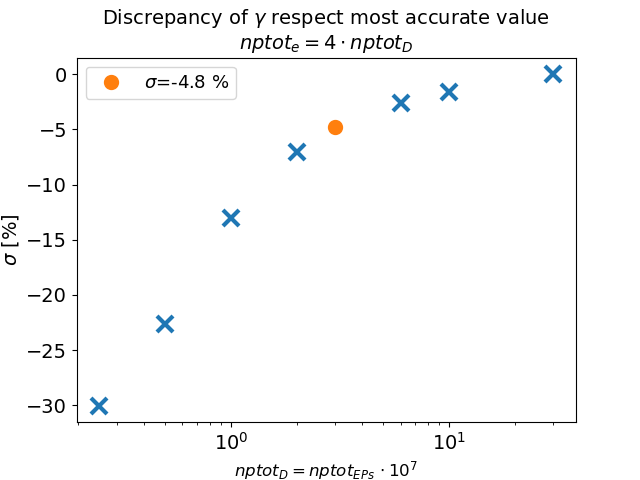}{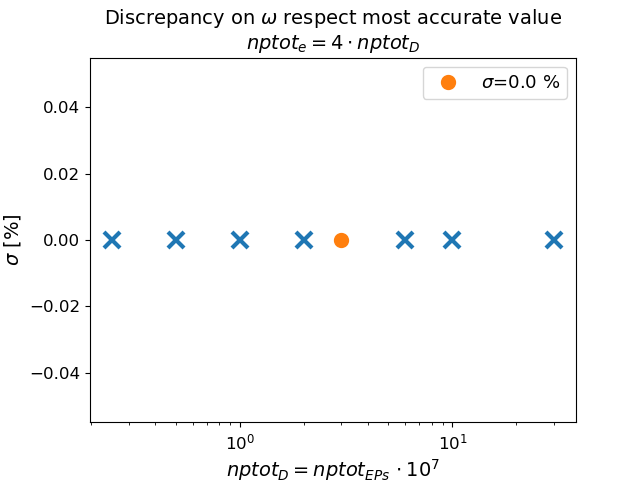}{Simulations with $\Delta t=4\,\omega_{ci}^{-1}$ and $N_{r}=1000$. Scan against the $\#$ of markers of Deuterium and EPs. The number of markers for the electrons is four times higher than that of the EPs. In the plots the discrepancies $\sigma$ of growth rate (\textbf{left}) and frequencies (\textbf{right}) with respect to the best resolved case (corresponding to the simulation with number of markers $=30\cdot 10^{7}$) are shown. The orange points correspond to the results obtained with the parameters shown in Tab.~\ref{TabC6:ref_sims}.}{FigC6:scan_nptot}
 
 \yFigTwo{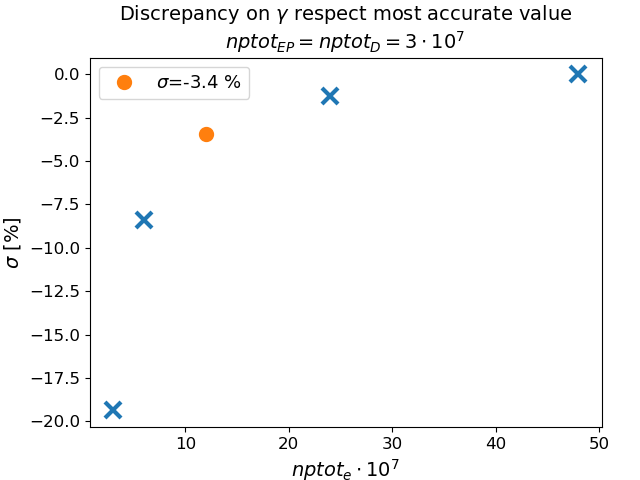}{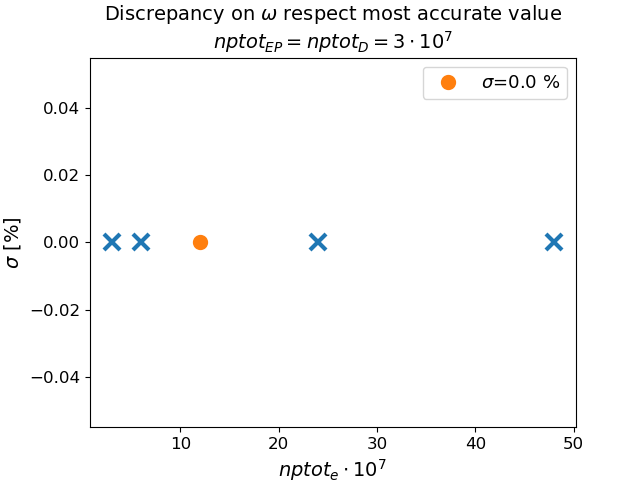}{Simulations with $\Delta t=4\,\omega_{ci}^{-1}$ and $N_{r}=1000$. Scan against the $\#$ of markers of the electrons. The number of markers for EPs and Deuterium particles is fixed at $3\cdot 10^{7}$. In the plots the discrepancies $\sigma$ of growth rate (\textbf{left}) and frequencies (\textbf{right}) with respect to the best resolved case (here corresponding to the simulation with number of markers $=50\cdot 10^{7}$) are shown. The orange points correspond to the results obtained with the parameters shown in Tab.~\ref{TabC6:ref_sims}. }{FigC6:scan_nptote}
 
  \yFigTwo{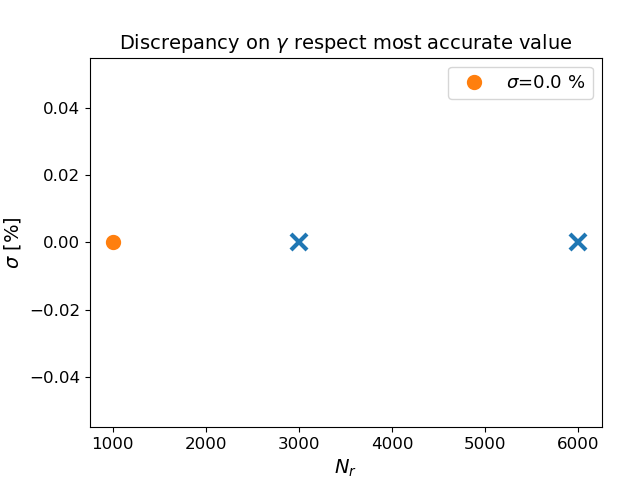}{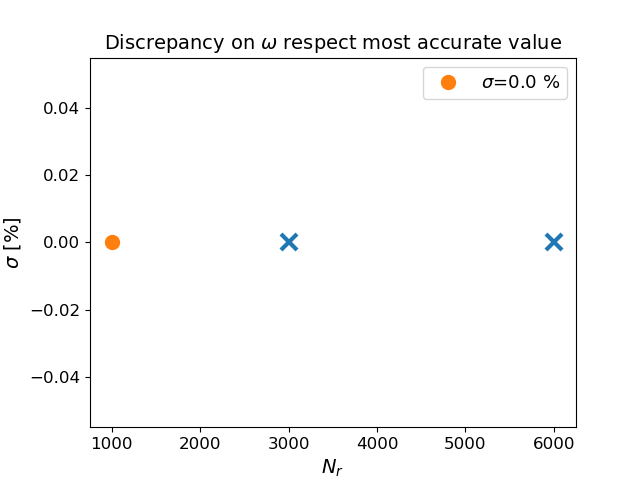}{Simulations with $\Delta t=4\,\omega_{ci}^{-1}$ and $\#$ of markers$_{D,e,EP}=(3,12,3)\cdot 10^{7}$. Scan against the $\#$ of radial grid points $N_{r}$. In the plots the discrepancies $\sigma$ of growth rate (\textbf{left}) and frequencies (\textbf{right}) with respect to the best resolved case (here corresponding to the simulation with $N_{r}=6000$) are shown. The orange points correspond to the results obtained with the parameters shown in Tab.~\ref{TabC6:ref_sims}. }{FigC6:scan_Ns}

\clearpage
\bibliographystyle{unsrturl}
\bibliography{main}
\end{document}